\documentclass[12pt,preprint]{aastex}
\usepackage{multirow}

\newcommand\arcpt{${{\lower3pt\hbox{$^{\prime\prime}$}}\atop{\raise4pt\hbox{.}}}$}

\slugcomment{submitted to the {\it Astronomical Journal} 15 JUNE 2010}

\shorttitle{64 system parallaxes from the CTIOPI program}
\shortauthors{Riedel et al.}

\begin{document}

\title{The Solar Neighborhood. XXII. Parallax Results from the CTIOPI
  0.9m Program: Trigonometric Parallaxes of 64 Nearby Systems
  with 0\farcs5 $\leq \mu \leq$ 1\farcs0 yr$^{-1}$ (SLOWMO sample)}

\author{Adric~R.~Riedel\altaffilmark{1}}

\affil {Department of Physics and Astronomy, Georgia State University,
Atlanta, GA 30302-4106} 

\email{riedel@chara.gsu.edu}

\author {John~P.~Subasavage\altaffilmark{1}}

\affil {Cerro Tololo Inter-american Observatory, La Serena, Chile}

\email {jsubasavage@ctio.noao.edu}

\author {Charlie~T.~Finch\altaffilmark{1}}

\affil {Astrometry Department, U.S. Naval Observatory, Washington DC
20392}

\email {finch@usno.navy.mil}

\author{Wei~Chun~Jao\altaffilmark{1}, Todd~J.~Henry\altaffilmark{1},
  Jennifer~G.~Winters\altaffilmark{1}, Misty~A.~Brown\altaffilmark{1}}

\affil {Department of Physics and Astronomy, Georgia State University,
Atlanta, GA 30302-4106}

\email{jao@chara.gsu.edu, thenry@chara.gsu.edu, winters@chara.gsu.edu,
  brown@chara.gsu.edu}

\author{Philip A.~Ianna\altaffilmark{1}}

\affil {Department of Astronomy, University of Virginia,
  Charlottesville, VA 22904}

\email{pai@virginia.edu}

\author{Edgardo Costa\altaffilmark{1}, Rene A. Mendez\altaffilmark{1}}

\affil {Departamento de Astronomia, Universidad de Chile, Santiago,
  Chile}

\email{costa@das.uchile.cl, rmendez@das.uchile.cl}

\altaffiltext{1}{Visiting Astronomer, Cerro Tololo Inter-american
  Observatory.  CTIO is operated by AURA, Inc.\ under contract to the
  National Science Foundation.}

\begin{abstract}

We present trigonometric parallaxes of 64 stellar systems with proper
motions between 0\farcs5 yr$^{-1}$ and 1\farcs0 yr$^{-1}$ from the
ongoing RECONS (Research Consortium On Nearby Stars) parallax program
at CTIO (the Cerro Tololo Interamerican Observatory).  All of the
systems are south of DEC $= +30$, and 58 had no previous trigonometric
parallaxes.  In addition to parallaxes for the systems, we present
proper motions, Johnson-Kron-Cousins $VRI$ photometry, variability
measurements, and spectral types.  Nine of the systems are multiple;
we present results for their components, three of which are new
astrometric detections. Of the 64 systems, 56 are within 25 parsecs of
the Sun and 52 of those are in the southern hemisphere, comprising
5.7\% of the total number of known southern 25 parsec systems.

\keywords{stars:standards --- stars:astrometry --- stars:astrometry
--- stars:parallax --- stars:M dwarfs}

\end{abstract}

\section{Introduction}
\label{sec:intro}

One of the most important products of nearby star research will be a
volume-limited sample of nearby stars.  All-inclusive volume-limited
samples of stars will allow us to answer large numbers of questions
about the formation, kinematics, stellar mass fraction, and
metallicity distributions of stars and (if representative) the galaxy
itself.  Due to the faintness of the M dwarfs that make up \emph{at
least} 72\% of all nearby stars, surveys aiming at completeness are
easiest carried out near the Sun.  Large numbers of potential nearby
stars have been found in ongoing series of proper motion surveys, most
notably (among others too numerous to mention here) the surveys of
Luyten \citep{1979lcse.book.....L}, Giclas
\citep{1958LowOB...4....1G}, Pokorny \citep{2003A&A...397..575P},
ourselves \citep{2005AJ....130.1658S}, and Lepine
\citep{2005AJ....130.1680L}.  Various other surveys \citep[for
instance,][]{2004A&A...421..643R,2008AJ....136.1290R,2008A&A...484..575J}
have followed up on those discoveries with spectroscopy and photometry
to determine distances and spectral properties of these systems,
although a great many still remain uninvestigated.  For our purposes,
however, we need accurate distances to prove membership, obtain
accurate luminosities, and good kinematics.

While there are many methods used to determine stellar distances, the
most accurate method remains the trigonometric parallax. It does not
rely on any prior knowledge about the star (as photometric and
spectroscopic ``parallaxes'' do), nor does it require a cluster of
stars (as secular and statistical parallax do), nor a companion (as
orbital parallax does). Trigonometric parallax is a geometric process,
involving only the Earth's orbital motion and the distance to the
star, operating independently of any unusual or misleading properties
a star might have.

\begin{figure}
\center
\includegraphics[angle=90,scale=.4]{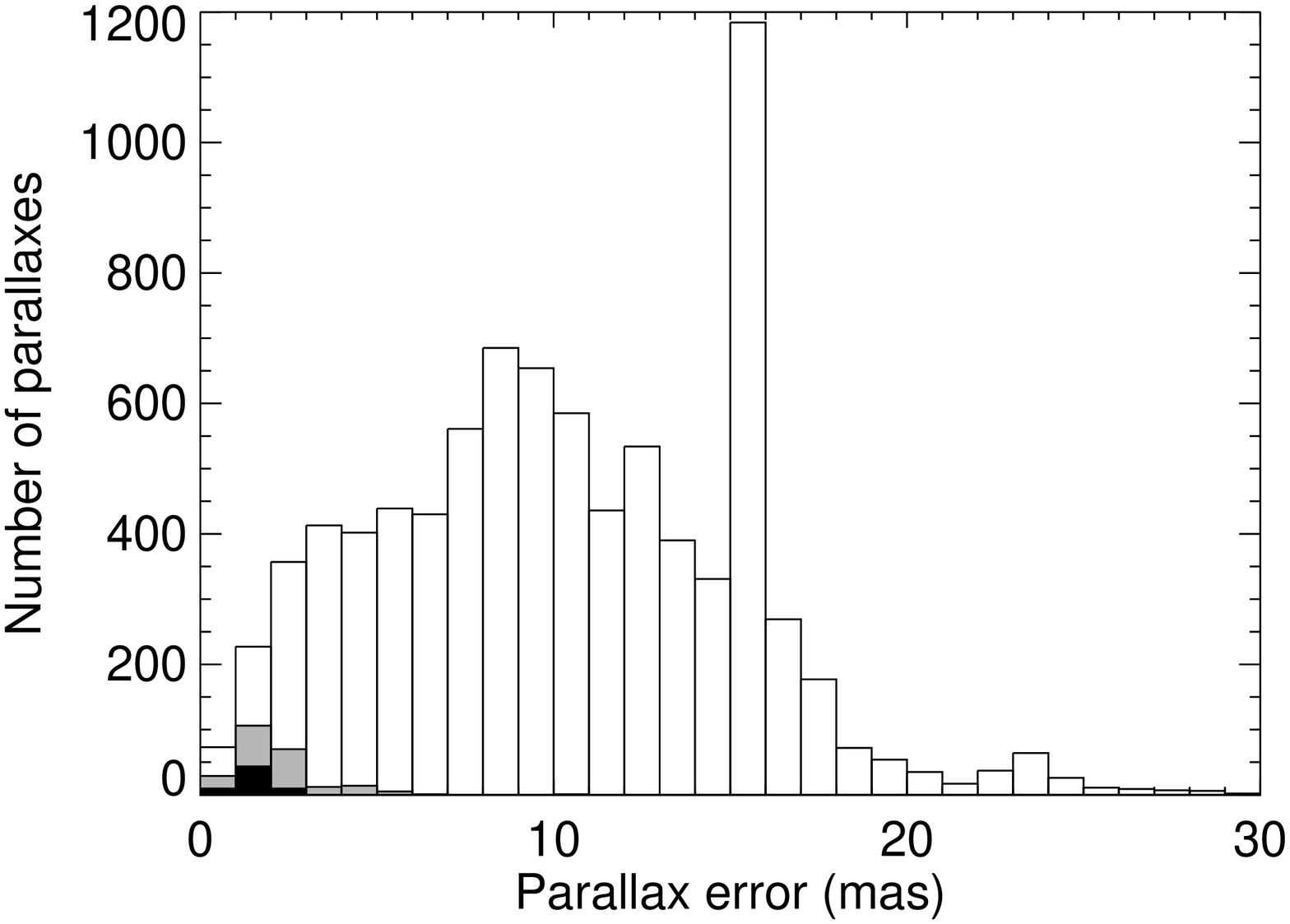}
\caption{Parallax errors for systems in this paper (black), and
  previous CTIOPI parallax papers (gray) are shown versus previous
  ground-based parallaxes from YPC (white, extends off this
  graph). Our improved precision is due to our CCD-based astrometry
  while the bulk of previous work was done with photographic plates.
  A few systems in this paper are also in YPC (see Table
  \ref{tab:weightedmeans}).  The enormous spike at 15 mas is a
  result of the methods used in YPC to assign errors to parallaxes
  published without error.
  \label{fig:precision}}
\end{figure}

The primary compendia of trigonometric parallaxes are the General
Catalog of Trigonometric Stellar Parallaxes, Fourth Edition
\citep{1995gcts.book.....V}, more commonly known as the Yale Parallax
Catalog (YPC), and the {\it Hipparcos} mission
\citep{1997A&A...323L..49P}.  The YPC is a compilation of 15994
parallax measurements for 8112 stars from various observatories,
representing ground-based parallax efforts prior to 1995 for stars as
faint as $V = 21.5$.  The parallaxes have a broad range of errors, but
most are between 2 and 20 milliarcseconds (mas), as shown in Figure
\ref{fig:precision}.  The latest version of the {\it Hipparcos}
mission's catalog \citep{2007A&A...474..653V} contains 117955 stellar
parallaxes, generally with errors less than 1 mas for stars brighter
than $V = 8$.  These compendia have already enabled statistical
studies of stellar formation, evolution, composition, kinematics, and
populations, making detailed large-scale surveys like the
Geneva-Copenhagen Survey \citep{2009A&A...501..941H} of F and G
dwarfs, for example, finally possible.

Despite the size of its catalog, {\it Hipparcos} did not find the
complete population of nearby stars.  The faint magnitude limit of
{\it Hipparcos} was $V \sim 13$, with a completeness limit of $V \sim
7.3$ \citep{1997A&A...323L..49P}.  Populations of stellar types AFGK
within 100 pc were richly sampled by {\it Hipparcos}, but
intrinsically faint stars such as M dwarfs, cool subdwarfs, and white
dwarfs largely remain the purview of the YPC, or of new ground-based
efforts such as the one discussed here.  A truly inclusive,
volume-limited survey of all types of stars on the scale of the
Geneva-Copenhagen survey is still not possible --- large numbers of
stars in the solar neighborhood are still unconfirmed, large numbers
of M dwarfs are still missing from nearby star
lists. \footnote{updates at http://www.recons.org/census.posted.htm}

The mission of RECONS (Research Consortium On Nearby
Stars)\footnote{http://www.recons.org} is to understand the solar
neighborhood, including the discovery, confirmation, and
characterization of nearby stars and their environments.  RECONS has
been operating a trigonometric parallax program called CTIOPI (Cerro
Tololo Interamerican Observatory Parallax Investigation) at the CTIO
0.9m telescope since 1999 with a primary goal of pushing the
parallax-verified solar neighborhood sample --- systems within 25 pc,
with a concentration on those within 10 pc --- toward completion.
This is the fifth paper of results from the ongoing program at the
0.9m, following \citet{2005AJ....129.1954J},
\citet{2006AJ....132.2360H}, \citet{2007ApJ...669L..45G}, and
\citet{2009AJ....137.4547S}; more details of the program can be found
in the preceding references.

While this paper does not finish off the sample of all stars in this
proper motion range, it furthers the CTIOPI goal of completing the
census of the solar neighborhood.  It contains 67 new trigonometric
parallaxes of 64 systems with proper motions between 0\farcs5
yr$^{-1}$ and 1\farcs0 yr$^{-1}$, our ``SLOWMO'' sample.  Of the 64
systems, 56 are within 25 pc of the Sun, and all are south of DEC$ =
+30$.  In addition to the parallaxes, we provide new measurements of
proper motions, Johnson-Kron-Cousins $VRI$ photometry, variability,
spectral types, and astrometric measurements of multiple systems.

\section{The Sample}
\label{sec:sample}

Completing the stellar census within 25 pc is a daunting task.  If we
assume the current census of systems within 5 pc is accurate (50
systems,\footnote{http://www.recons.org/TOP100.posted.htm} see
\citet{2006AJ....132.2360H} for the most recent additions within five
pc) and representative of the stellar space density of the solar
neighborhood, we expect 6250 star systems within 25 pc.  The most
complete parallax-selected list of objects remains the NStars
Database, which includes only 2011 systems within 25 pc
\citep{2002AJ....123.2002H}, indicating that the sample is only 32\%
complete.  Clearly, much more work is needed to achieve a truly
comprehensive volume-limited sample of space; only then will we truly
be able to characterize the compositional nature of the Galaxy. Even
the Gaia mission, with a limiting magnitude of $V = 20$, will not
reach the end of the main sequence past ten parsecs.

The 64 systems discussed in this paper are listed in Tables
\ref{tab:astrometry} and \ref{tab:photometry}.  They were selected for
the CTIOPI program for a variety of reasons: either their high proper
motion made them targets for M. Brown's masters thesis on SLOWMO
systems, their estimated distances suggested they might be within 10
parsecs, or they had YPC parallaxes with large errors that placed the
system within 10 parsecs. The systems themselves are all from
\citet{1979lcse.book.....L}, \citet{1957QB6.L98........},
\citet{1991A&AS...91..129W}, \citet{1994A&AS..105..179W},
\citet{2000A&A...353..958S}, and a private communications with
\citet{1999Scholz} for APMPM J2127-3844.  Most of them were
investigated for companions in \citet{2003AJ....125..332J}.  All of
the systems have proper motions of 0\farcs5--1\farcs0 yr$^{-1}$, have
red dwarf primaries with $V =$ 10.35--19.17 and have spectral types
M1.0V to M6.0V.  Seven of the systems presented here have known or
suspected companions; we confirm six of them and present individual
parallax measurements for components of three.  We have also
discovered evidence of multiplicity for a further three systems and
suspect additional components in five more systems; thus 14\% (9 out
of 64) of our systems are multiple, and 9\% (5 out of 64) are
suspected multiples.

This sample builds upon our previous efforts that also revealed
systems within 25 pc, including stars with $\mu \geq$ 1\farcs0
yr$^{-1}$ \citep{2005AJ....129.1954J} (our MOTION sample), red dwarfs
within 10 pc \citep{2006AJ....132.2360H}, and white dwarfs within 25
pc \citep{2009AJ....137.4547S} from the CTIOPI 0.9m program, and mixed
samples from our 1.5m program
\citep{2005AJ....130..337C,2006AJ....132.1234C}.

\section{Observations}
\subsection{Astrometric and Photometric Observations}
\label{sec:astrometryobs}

All astrometric and photometric observations were carried out at the
CTIO 0.9m telescope, initially (1999-2003) under the aegis of the NOAO
(National Optical Astronomy Observatory) Surveys Program, and later
(2003-present) via the SMARTS (Small and Moderate Aperture Research
Telescope System) Consortium.  The observations presented in this
paper were obtained between 1999 and 2009 utilizing the center
1024x1024 pixels of the 0.9m telescope's Tek 2048x2046 CCD and CTIO's
$V_J, R_{KC}$, and $I_{KC}$ (hereafter without
subscripts)\footnote{Subscripts: ``J'' indicates Johnson, and ``KC''
indicates Kron-Cousins.  The central wavelengths for $V_J, R_{KC}$,
and $I_{KC}$ are 5475, 6425, and 8075 \AA, respectively.}  filters.
Additional details of the observing protocol can be found in
\citet{2005AJ....129.1954J}.

Four significant instrumental events in the course of the CTIOPI
program have affected data published in this paper:
\begin{itemize}

\item In February 2005, the Tektronix \#2 $V$ filter (hereafter ``old
  $V$'') used by CTIOPI cracked and was replaced by the
  almost-photometrically identical (transmission properties and
  bandwidth) Tektronix \#1 $V$ filter (hereafter ``new $V$''). With
  four years of new $V$ data, we are able to make some comparisons
  between the two:

\subitem{As reported in \citet{2009AJ....137.4547S}, the new $V$
  filter is photometrically consistent with the old $V$ filter to
  within reported CTIOPI accuracies \citep[0.03
  mag,][]{2004AJ....128.2460H}, although we find our V filter
  photometry is only accurate to 0.05 mag in this dataset.}

\subitem{Also as reported in \citet{2009AJ....137.4547S}, some new $V$
  filter data cause a few-mas offset in the RA axis astrometric
  residuals.  This is endemic to the new $V$ filter itself and is not
  the result of changing filters; recent data taken with the old $V$
  filter show no such behavior in the residuals when added to older
  data. Parallax results using only new $V$ filter data are slightly
  but non-systematically displaced relative to results using only old
  $V$ data, which were found to be consistent with YPC and {\it
  Hipparcos} parallaxes in \citet{2005AJ....129.1954J}.  Part of the
  error appears to depend on the filter itself, the rest appears to
  depend on coverage: tests were only conducted on stars with large
  datasets before and after the $V$ filter replacement. Those stars
  tend to have better coverage of the parallax ellipse in old $V$
  (earlier) than in new $V$ (later, when their parallaxes were well
  determined and they became lower priority targets).  Even so, the
  new $V$ parallax is usually within 2-$\sigma$ of the old $V$
  measurement, and mixed $V$ parallaxes are always within
  2-$\sigma$. All parallaxes in this paper using new $V$ filter data
  are noted in Table \ref{tab:astrometry}.}

\item In April 2005, the Telescope Control System (TCS) on the 0.9m was
  completely replaced and refurbished, yielding improved pointing and
  tracking.  No astrometric effects have been detected in datasets
  spanning the TCS upgrade.

\item On 7 March 2009, a power outage damaged the gain $=$ 1 circuitry
  for the CCD, and CTIOPI began using gain $=$ 2. The differences
  between the two gains are purely electrical, and tests confirm that
  the switch does not affect our astrometry, as expected.

\item In July 2009, the old $V$ filter was returned to service.
  Extensive tests showed the hairline crack near the edge does not
  affect data acquired on the central quarter of the CCD as used in
  CTIOPI.  Furthermore, testing many stellar fields indicated no
  adverse effects on the astrometry when reducing data with and
  without recent data in the old $V$ filter.  Recent old $V$ filter
  data have been used in this paper's astrometric reductions of LHS
  1050, LHS 1561, LHS 3443, LHS 4009AB, and LHS 4016.

\end{itemize}

\subsection{Spectroscopic Observations}
\label{sec:spectroscopyobs}

We carried out spectroscopic observations at two telescopes to
determine the spectral types of 25 of the objects listed in Table
\ref{tab:photometry}.  From 2003--2006, we used the CTIO 1.5m, the R-C
Spectrograph with a Loral 1200 $\times$ 800 CCD, and the 32/I grating
to obtain spectra covering 6000--9500\AA~at a resolution of 8.6\AA.
For WT 244 and GJ 438, we used the CTIO 4m, the R-C spectrograph with
a Loral 3X $\times$ 1K CCD, and the 181 grating to obtain spectra
covering 5500--10000\AA~at a resolution of 5.7\AA.  Further details
concerning the 1.5m spectroscopy program and associated data reduction
can be found in \citet {2004AJ....128.2460H}, while details of the 4m
spectroscopy program can be found in \citet{2002AJ....123.2002H}.

\section{Results}
\subsection{Astrometry --- Parallaxes and Proper Motions}
\label{sec:pi}

Parallaxes and proper motions of 64 stellar systems are presented in
Table \ref{tab:astrometry}.  Nine of our systems have multiple
parallax measurements, from YPC, our CTIOPI 1.5m program, or multiple
components published in this paper.  For these cases, we present
weighted average system parallaxes in Table \ref{tab:weightedmeans}.

All parallax data were analyzed with the custom IRAF/IDL pipeline used
in CTIOPI publications since 2005, using an iterative Gaussfit program
described in \citet{2005AJ....129.1954J}.  Starting in 2007, the
reduction methodology was changed by the implementation of a new
SExtractor centroiding algorithm, described in
\citet{2009AJ....137.4547S}.

As always, CTIOPI parallaxes must meet several criteria before they
are deemed fit to publish.  First, to ensure good coverage of the
parallax ellipse, there is an informal limit of 30 frames each in the
`morning' and `evening' halves of the ellipse with a goal of having 20
usable frames each; the number of frames used in the final reductions
ranged from 45 (LHS 2899) to 137 (LHS 1630AB).  Second, the system
must be followed for about two years to decouple the star's motion
into parallax and proper motion; the coverage in this paper varies
from 1.99 years (LHS 2335) to 10.15 years (LHS 4009AB).  Third, CTIOPI
targets are expected to have parallax errors less than 3 mas before
publishing.  The smallest parallax error we are publishing here is
0.63 mas (GJ 1157) and the largest is 2.57 mas (LHS 1050 and LHS
2122), while the median error on these parallaxes is 1.37 mas.

Our parallaxes are derived by measuring the motion of the target star
(pi star) relative to background reference stars, and then correcting
the parallactic motion to an absolute reference frame with $VRI$ band
photometric distance estimates to the reference star ensemble (not to
be confused with our $VRIJHK$ distance estimates for red dwarfs,
discussed in $\S$\ref{sec:photdistances}).  Good reference star fields
consist of at least five but rarely more than twelve stars that are as
close as possible to the pi star on the CCD, have more than 1000 peak
counts, and surround the pi star as much as possible.  The most
accurate parallaxes are obtained with exposures longer than 90
seconds, which is long enough at CTIO for images in the $VRI$ bands to
smooth over atmospheric effects and consequently improve centroids.
We have also re-observed several systems (e.g. LHS 1561, LHS 3909, and
LHS 3443) after a multi-year hiatus to improve proper motions and
possibly reveal long-period perturbations (see
$\S$\ref{sec:perturbations}), otherwise the extra observations have a
minor effect on the derived parallax.

\subsection{Astrometry --- Multiples}
\label{sec:multiples}

Seven of our systems contain known multiples: LHS 1630AB, LHS 1749AB,
LHS 1955AB, LHS 2567/2568, LHS 3001/3002, LHS 3739/3738AB, and LHS
4009AB.  We have resolved orbital motion above the 3-$\sigma$ level
(angular motion or changes in separation) for five of those systems,
as described in Table \ref{tab:multiples}. Apart from LHS 1955AB, all
values published in Table \ref{tab:multiples} are derived from at
least three frames on the nights listed.

The components of LHS 1630AB, LHS 1749AB and LHS 1955AB are close
enough that their PSFs (point spread functions) overlap.  In the case
of LHS 1630AB the B component was never fully resolved, but it does
appear as an elongation to the PSF. LHS 1749AB was only resolved on 15
frames from four nights. LHS 1955AB was only resolved on seven frames
from five nights using restricted centroiding parameters that enabled
the separation of blended sources; the two frames from the earliest
night and the one frame from the latest night are used to derive the
results in Table \ref{tab:multiples}.

Errors presented in Table \ref{tab:multiples} include both measurement
and systematic errors. The systematic errors were computed from the
nights of data measured for Table \ref{tab:multiples}, with the
exception of the three frames used for LHS 1955AB.  All frames from a
single visit (one night of observations on one system) were compared
to the 2MASS All-Sky Point Source Catalog using {\it imwcs}
\footnote{A World Coordinate System setting program from the WCSTools
library, http://tdc-www.harvard.edu/software/wcstools/}; the standard
deviations of the plate scales and rotations per-visit were then
averaged across all visits to get a more representative error.
Systematics for the CTIO 0.9m on those frames give a 0.015\% error in
the plate scale (and therefore separations), and a 0.0083 degree error
in the rotation (and therefore position angles).  In all cases, the
measurement errors dominate the systematic errors.

\subsection{Astrometry --- New astrometric multiples}
\label{sec:perturbations}

\begin{figure}
\center
\includegraphics[angle=90,scale=.4]{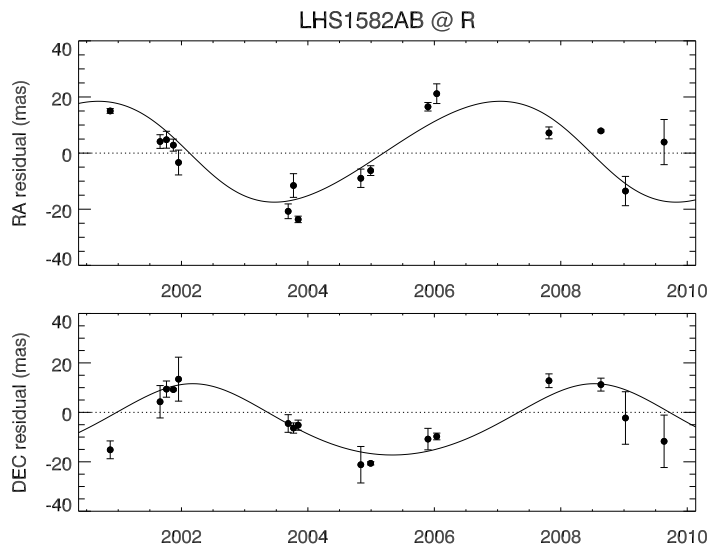}
\caption{Plots of the nightly means of our astrometric residuals in RA
  and DEC for LHS 1582AB after solving for parallax and proper motion.
  A perturbation with a $\sim$6 year period is evident, and the
  resultant orbital fit (Table \ref{tab:orbits}) is plotted on this
  graph.  Two nights with only a single CCD image each (obtained for
  photometry) are not shown or used in the orbital solution.
  \label{fig:lhs1582ab}}
\end{figure}

\begin{figure}
\center
\includegraphics[angle=90,scale=.4]{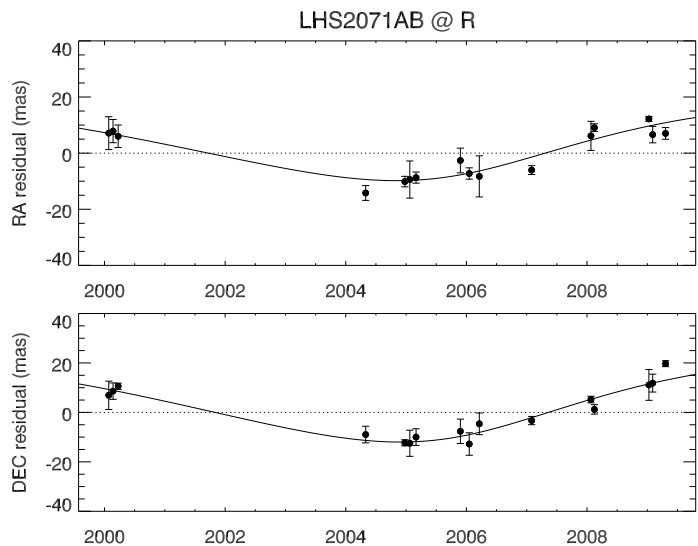}
\caption{Plots of the nightly means of our astrometric residuals in RA
  and DEC for LHS 2071AB after solving for parallax and proper motion.
  A perturbation is evident, but the orbit has not wrapped in our data
  so the plotted orbital fit (Table \ref{tab:orbits}) is poorly
  determined. \label{fig:lhs2071ab}}
\end{figure}

\begin{figure}
\center
\includegraphics[angle=90,scale=.4]{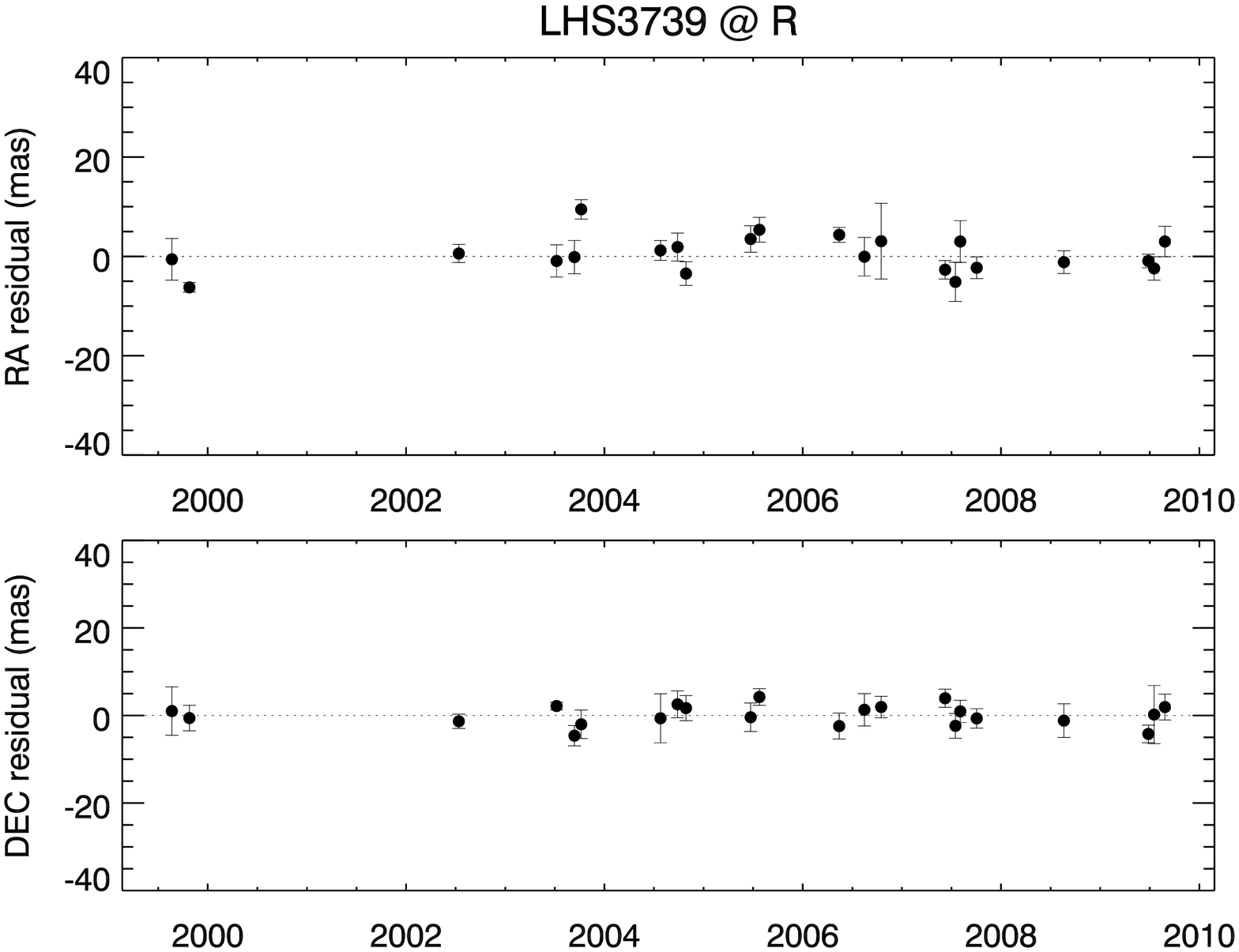}
\includegraphics[angle=90,scale=.4]{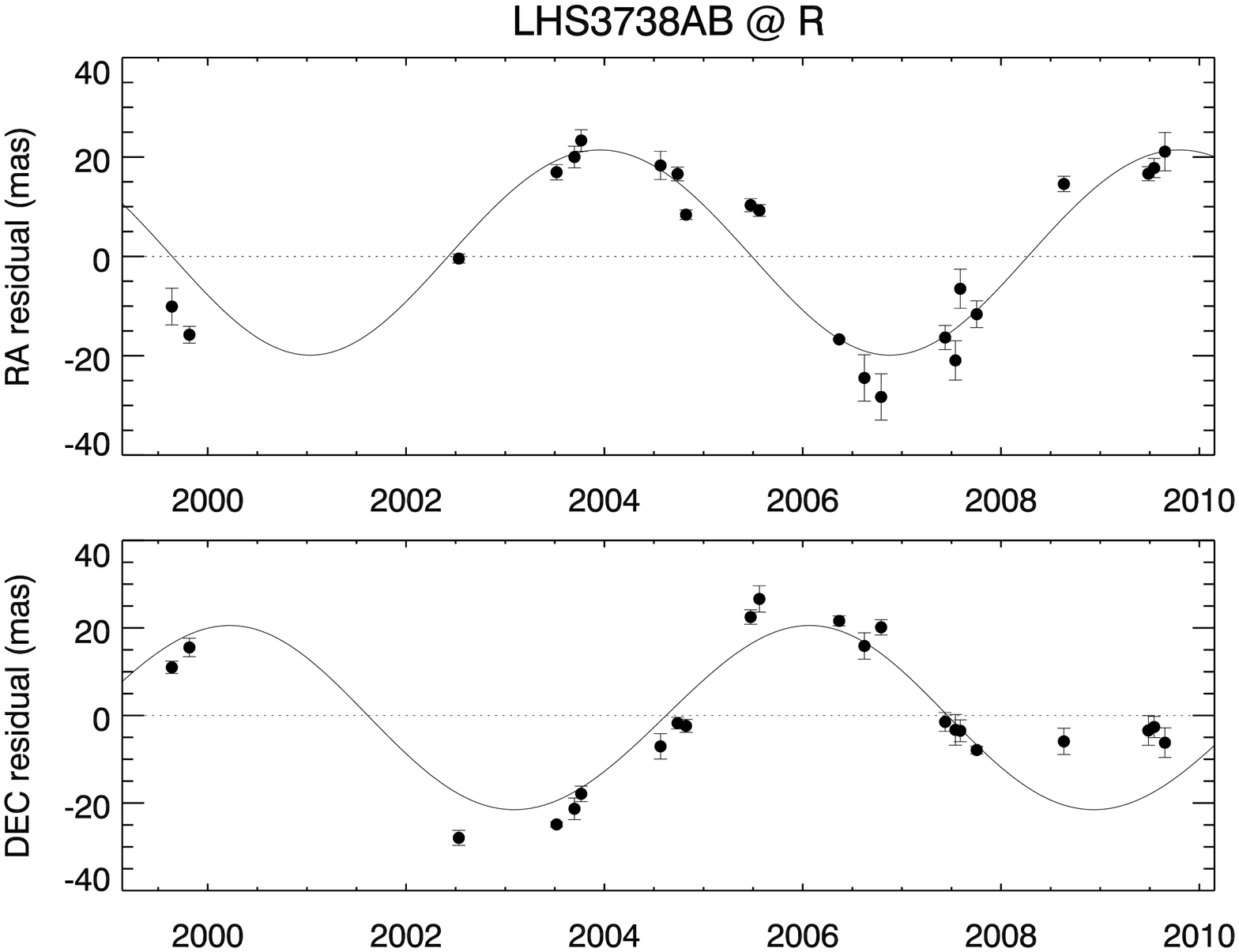}
\caption{Plots of the nightly means of our astrometric residuals in RA
  and DEC for LHS 3739 (top) and LHS 3738AB (bottom) after solving for
  parallax and proper motion.  LHS 3738AB clearly shows a perturbation
  with a period of $\sim$6 years whose orbital fit (Table
  \ref{tab:orbits}) is overplotted; using the same CCD frames and
  reference stars the residuals for LHS 3739 remain flat.  Two nights
  with only a single CCD image each (obtained for photometry) are not
  shown or used in the orbital solution. \label{fig:lhs3739}}
\end{figure}

\begin{figure}
\center
\includegraphics[angle=90,scale=.4]{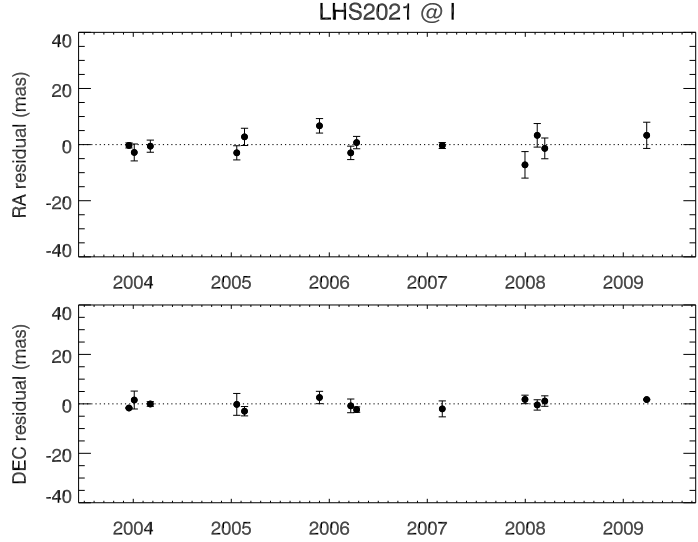}
\caption{Plots of the nightly means of our astrometric residuals in RA
  and DEC for LHS 2021 after solving for parallax and proper motion.
  This star appears to a single main-sequence M dwarf.  Two nights
  with only a single CCD image each (observed for photometry) are not
  shown. \label{fig:lhs2021}}
\end{figure}

\begin{figure}
\center
\includegraphics[angle=90,scale=.4]{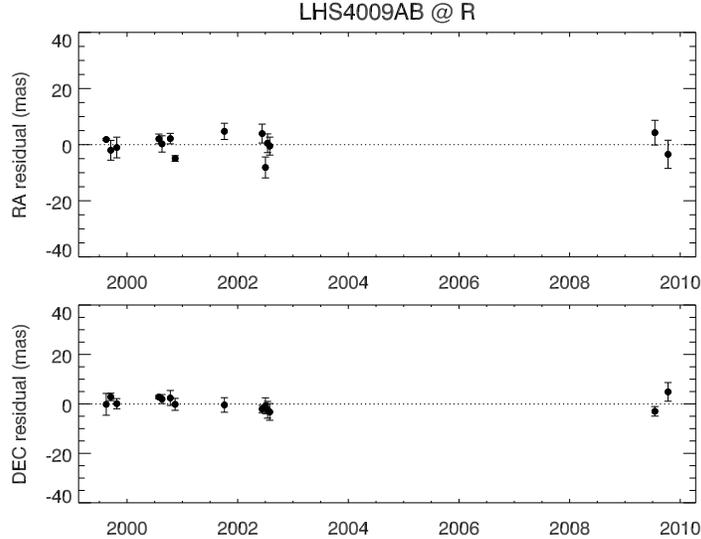}
\caption{Plots of the nightly means of our astrometric residuals in RA
  and DEC for LHS 4009AB after solving for parallax and proper motion.
  There is no perturbation evident.  The two components discovered in
  \citet{2006A&A...460L..19M} have a magnitude difference of only
  $\Delta K$=0.1 and an orbital period $\sim$3 years. We see no
  evidence of duplicity in the data.  One night with only a single
  CCD image (obtained for photometry) is not shown.
  \label{fig:lhs4009ab}}
\end{figure}

Three of the systems discussed here --- LHS 1582AB, LHS 2071AB, and
LHS 3738AB --- have been found to be previously undetected astrometric
binaries, as shown in Figures \ref{fig:lhs1582ab},
\ref{fig:lhs2071ab}, and \ref{fig:lhs3739}, respectively. For
comparison, three additional stars --- LHS 2021, LHS 3739 and LHS
4009AB --- are shown in Figures \ref{fig:lhs2021}, \ref{fig:lhs3739}
and \ref{fig:lhs4009ab}.  LHS 2021 and LHS 3739 appear to be single
stars while LHS 4009AB is a known close binary that also appears
single in our data.  We re-observed LHS 4009AB several years after the
parallax was finished to search for long-term perturbations; none was
found.  All five systems are discussed in detail in
$\S$\ref{sec:systemnotes}.

The CTIOPI parallax reduction pipeline fits the astrometric positions
of a star to a linear proper motion and a parallax ellipse of known
shape and unknown size.  Any further motion caused by the orbit of a
companion remains, appearing as a perturbation in the astrometric
residuals.  Unlike an orbit determined by a technique that resolves
two objects in a binary, an astrometric perturbation describes the
orbit of the photocenter, not the motion of any individual
component. (Astrometric perturbations are therefore greater for larger
magnitude differences between the components, larger companion masses,
and larger semi-major axes of the orbits; the photocenter of an
equal-mass equal-luminosity binary will not move at all.) We can still
solve for all orbital elements except semi-major axis.  In its place,
we can determine the semi-major axis of the photocentric orbit, which
can then be scaled to the relative orbit if the system is resolved.

Photocentric orbital elements for the three new astrometric binaries
were computed from the astrometric residuals using an iterative
Thiele-Innes least-squares solver \citep{1989AJ.....98.1014H} and are
given in Table \ref{tab:orbits}.  Points from nights with only a
single CCD image (generally obtained for the purpose of photometry)
were removed.  The orbits should be considered preliminary, as our
astrometric datasets do not have sufficient time coverage to publish
definitive orbits, particularly in the case of LHS 2071AB for which
the orbit is not complete.  LHS 2071AB and LHS 3738AB have now both
been resolved through followup work, as discussed in
$\S$\ref{sec:systemnotes}.

The parallaxes published in Table \ref{tab:astrometry} were computed
from data where the photocentric orbit was removed.  The orbital
position of the photocenter was calculated and subtracted from the
actual measured position of the photocenter at each data point,
including those from nights with only a single CCD image.  The
parallax was then re-reduced based on this new dataset.

CTIOPI's detection capabilities are limited by several factors.
Systems are typically observed 1--4 times a year and thus the data are
insensitive to periods less than a year.  The program has only been
running for ten years, and cannot wrap orbits with periods longer
than ten years.  CTIOPI also has a 3--6 mas nightly precision error
(depending on the specifics of the reference field) that limits our
sensitivity to low-amplitude binaries. Fortunately, astrometric
perturbations are typically self-confirming; genuine orbital motion
will show up in both the RA and DEC axes unless the orbit
is nearly north-south or east-west on the sky.  Nevertheless, to check
for systematics within the field we have reduced the three brightest
reference stars in each of our astrometric perturbation fields as if
they were the parallax target.  In only one case did a reference star
showed a perturbation of any kind; that reference was removed from the
reduction of LHS 2071AB.

\subsection{Photometry --- Variability}
\label{sec:variability}

The variability values in Table \ref{tab:photometry} are calculated
according to the methodology of \citet{1992PASP..104..435H} with
additional details given in \citet{2006AJ....132.2360H}. In Table
\ref{tab:photometry} we list the level of variability in magnitudes
(Column 8) of each target star in its parallax filter (Column 7).  The
number of nights on which each star was observed (Column 9) and the
number of frames (Column 10) used for the variability study are also
given.

Although many M dwarfs are minutely variable, none of the stars in
this sample have been found to vary by more than 2\% in the frame
series available.  The single exception is LHS 1749AB at 0.028 mag.
In this case, the variability is likely due to the B component at a
separation of 3\arcsec~falling within the relative photometry
aperture, and variations in seeing affecting the extracted fluxes.

\subsection{Photometry --- Standard}
\label{sec:photometry}

$VRI$ magnitudes are listed in Table \ref{tab:photometry}, along with
extracted $JHK_s$ magnitudes from the Two Micron All Sky Survey
Catalog of Point Sources \citep{2006AJ....131.1163S}.  The errors on
the $VRI$ magnitudes are less than 0.05 mag for $V$, and 0.03 mag for
$R$ and $I$, with few exceptions, most notably the faint stars LHS
2021 and LHS 3002.  All photometric observations were reduced via a
custom IRAF pipeline and transformed onto a the Johnson-Kron-Cousins
system through the use of photometric standards from
\citet{1992AJ....104..340L}, \citet{2007AJ....133.2502L} and
\citet{1982PASP...94..244G}, as described in
\citet{2006AJ....132.2360H}.

\subsection{Photometry --- Distance Estimates}
\label{sec:photdistances}

\begin{figure}
\center
\includegraphics[angle=90,scale=.4]{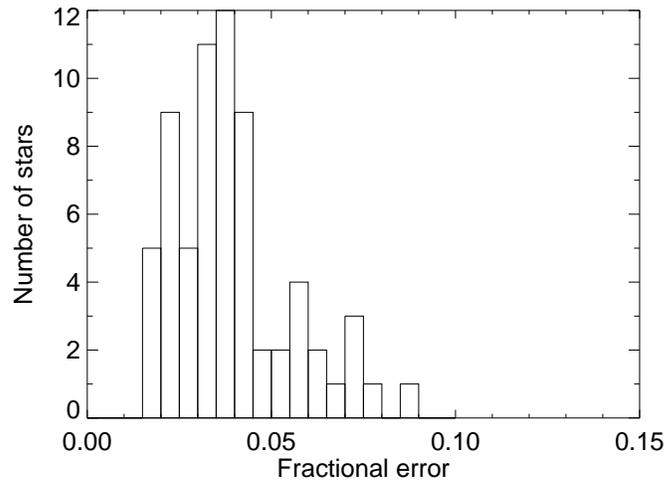}
\caption{The internal photometry distance errors are shown without the
  15.3\% external systematic error.  The average error is 3.9\%,
  indicating that the distance estimates are remarkably consistent for
  this sample. \label{fig:errors}}
\end{figure}

For purposes of initial target selection as well as for additional
analysis once trigonometric distances are determined, we used our CCD
photometry combined with 2MASS $JHK$ photometry to calculate the
photometric distances listed in Table \ref{tab:photometry} (Columns 16
and 17).  The fourth-order polynomial fits for twelve color-absolute
magnitude relations, the process by which the relations were
determined, and the calculation of internal and external errors are
described in detail in \citet{2004AJ....128.2460H}.  Our quoted
photometric distances have internal errors (defined as the standard
deviation of the distances estimated from the twelve relations) below
10\% and an additional external systematic error of 15.3\%.  The
distribution of internal errors is shown in Figure \ref{fig:errors};
the average error is 3.9\%, which is much smaller than the external
errors.  The errors listed for the photometric distances given in
Table \ref{tab:photometry} include both internal and external errors.

Because the fits used for the photometric distance estimates are
derived using main sequence stars, the estimates are only accurate
when the objects are single, main-sequence, M dwarfs.  For the most
part, these systems are indeed single, and 54 of the 64 systems (84\%)
fall within the 2-$\sigma$ range when their combined internal and
external errors are considered, as shown in Figure
\ref{fig:distances}.  Stars above the 2-$\sigma$ line in Figure
\ref{fig:distances} are likely to be underluminous subdwarfs, while
those below the 2-$\sigma$ line are presumably overluminous multiples.
There are no subdwarfs in this sample, (although LHS 1050, LHS 1807,
and LHS 3739/3738AB may be slightly metal-poor) but there are several
known close multiples with combined photometry, either previously
known (LHS 1630AB, LHS 1955AB, LHS 4009AB), or discovered by us (LHS
1582AB, LHS 2071AB, LHS 3738AB).  There is considerable scatter in
$M_V$ along the main sequence, visible in Figure \ref{fig:HRdiagram},
with up to two full magnitudes of spread for early to mid M dwarfs.
An equal magnitude binary will have a distance estimated to be 41\%
closer via photometry than is determined trigonometrically, but given
the spread in $M_V$ in the main sequence, only further work will
confirm or refute the multiplicity of suspected targets.

\subsection{Spectral Typing}
\label{sec:spectraltypes}

Spectral types are given in Table \ref{tab:photometry}, and come from
five sources that can be arranged into two broad groups. One group is
from RECONS spectroscopy, detailed in \citet{1991ApJS...77..417K},
\citet{1994AJ....108.1437H}, and \citet{2002AJ....123.2002H}. RECONS
spectroscopy was used to determine the spectral types of all stars not
taken from literature, and by \citet{1995AJ....109..797K} to classify
LHS 1807.

The remaining spectral types are from the Palomar/Michigan State
University Nearby Star Spectroscopic Survey (PMSU)
\citep{1995AJ....110.1838R,1996AJ....112.2799H} and related paper
\citet{2007AJ....133.2825R}, all of which use the same weighted
spectral indicies method linked to the spectral standards in
\citet{1991ApJS...77..417K}.  Where RECONS classifications were done
over a range of 6000--9000\AA~with an effective resolution of
5.7--8.6\AA~depending on the setup ($\S$\ref{sec:spectroscopyobs}),
the PMSU program used 6200--7500\AA~with resolution 1.8\AA.  In
practice our results differ from PMSU results only occasionally and
never more than half a subtype (see Table \ref{tab:spectra} for a
comparison).

\section{Systems Worthy of Note}
\label{sec:systemnotes}

\begin{figure}
\center
\includegraphics[angle=90,scale=.7]{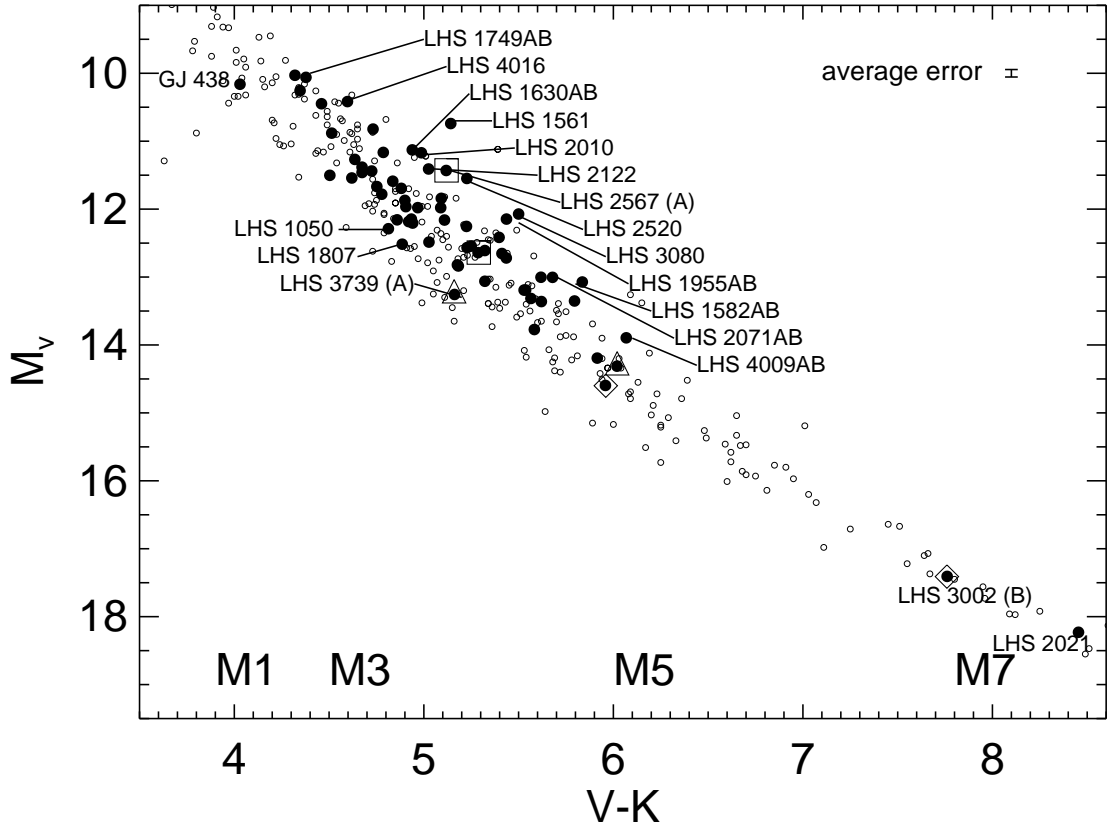}
\caption{All 67 system components with parallaxes reported here are
  plotted as large solid points on an observational HR diagram.  Small
  points represent stars in the RECONS 10 pc sample.  LHS 2567/2568
  are enclosed with squares, LHS 3001/3002 with diamonds, and LHS
  3739/3738AB with triangles.
  \label{fig:HRdiagram}}
\end{figure}

{\bf LHS 1050:} A 1.3-$\sigma$ underluminous single-star system
(11.7$\pm$0.3 pc trig/15.2$\pm$2.4 pc phot dist) that still appears to
be on the main sequence as shown in Figure \ref{fig:HRdiagram}.  The
YPC distance 11.5$\pm$1.8 pc is consistent with our distance,
11.7$\pm$0.3 pc. A weighted mean parallax is given in Table
\ref{tab:weightedmeans}.

\begin{figure}
\center
\includegraphics[angle=90,scale=.8]{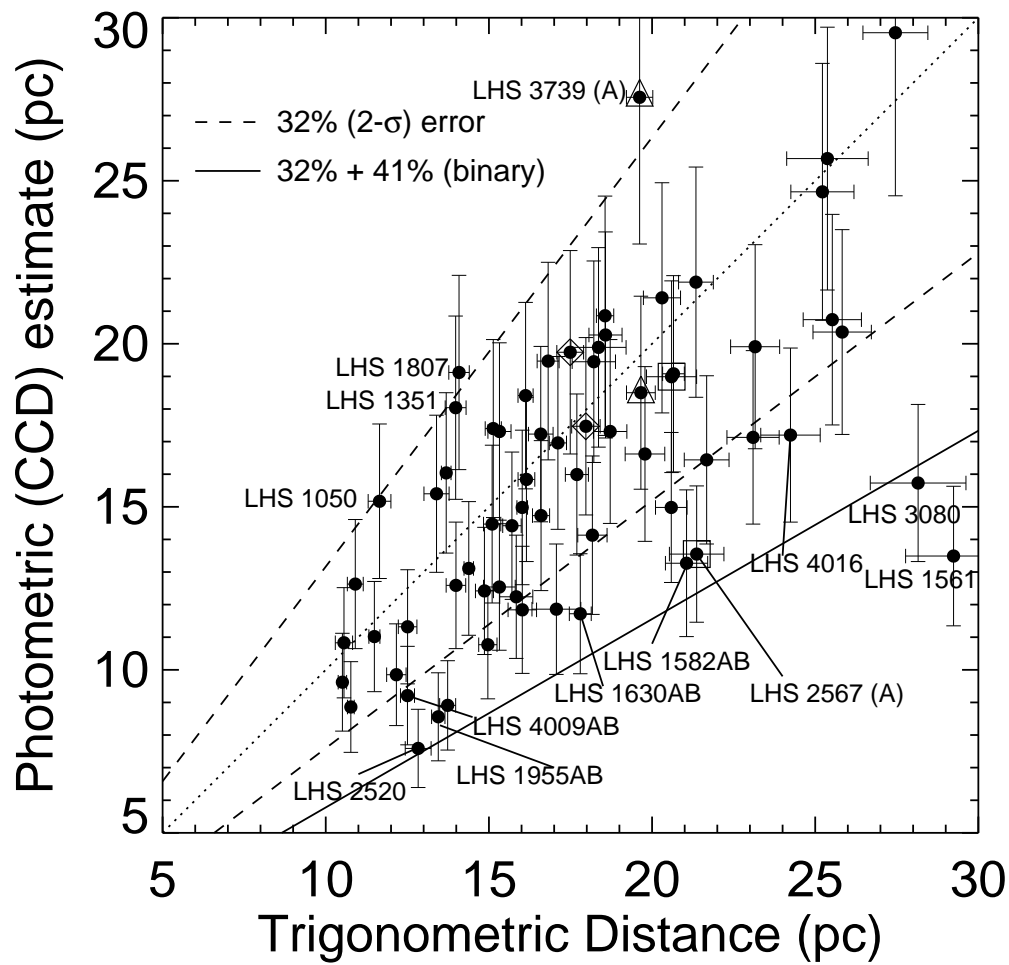}
\caption{Photometric distance estimates compared to trigonometric
  parallax distances, identical distances are plotted with a dotted
  line.  Dashed lines display the average 2-$\sigma$ error of our
  photometric distance estimates.  Beyond the solid line, even an
  equal-luminosity binary cannot fully account for the mismatch
  between trigonometric and photometric distance estimates.  LHS
  2567/2568 are plotted with squares, LHS 3001 (the nearer one by
  trigonometric parallax)/3002 with diamonds, and LHS 3739/3738AB with
  triangles.  LTT 5066 at 46 pc is not plotted.
  \label{fig:distances}}
\end{figure}

{\bf LHS 1561:} The most overluminous system in the sample, it has a
4.4-$\sigma$ distance mismatch (29.2$\pm$1.5 pc trig/13.5$\pm$2.1 pc
phot dist, see Figure \ref{fig:distances}), and is noticeably elevated
above the main sequence in Figure \ref{fig:HRdiagram}.  We see no
astrometric perturbation; it may be a multiple CTIOPI is not sensitive
to or a pre-main-sequence object.  The parallax has not `stabilized':
additional data continue to change the answer by more than 1-$\sigma$,
which is often a sign of unresolved orbital motion.

{\bf LHS 1582AB:} A new astrometric binary with a 6.4 yr period and an
18 mas photocentric semi-major axis (see Figure \ref{fig:lhs1582ab}).
It has a 2.7-$\sigma$ distance mismatch (21.1$\pm$0.7 pc/13.3$\pm$2.3
pc phot dist, see Figure \ref{fig:distances}) and is elevated above
the main sequence in Figure \ref{fig:HRdiagram}.  A preliminary orbit
is given in Table \ref{tab:orbits}; the orbital motion was removed
from the data before fitting the final parallax.

\begin{figure}
\center
\includegraphics[scale=.6]{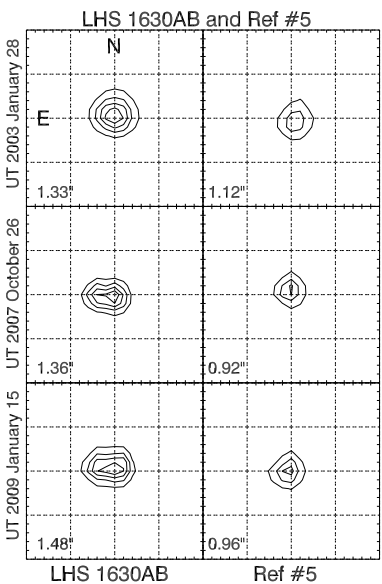}
\caption{Contour plots of LHS 1630AB (left) and example single star
  Ref \#5 (right, contours exaggerated 20 times) on three different
  nights in the $I$ filter.  LHS 1630B was first reported by
  \citet{2004A&A...425..997B} at separation 0\farcs61, angle 72 deg in
  2002. Grid markings are 5 pixels (2\farcs05), FWHM values for the
  PSFs are given in the lower left of each panel. \label{fig:lhs1630ab}}
\end{figure}

{\bf LHS 1630AB:} We confirm the Adaptive Optics companion reported in
\citet{2004A&A...425..997B} seen on 18 September 2002 with a
separation of 0\farcs61 at a position angle of 72 deg.  The B
component is visible in $I$ band photometry frames as of 2007, as
shown in Figure \ref{fig:lhs1630ab}.  The system has a 2.8-$\sigma$
distance mismatch (17.8$\pm$0.3 pc trig/11.7$\pm$1.8 pc phot dist) and
is elevated above the main sequence as seen in Figure
\ref{fig:HRdiagram}, but we see no astrometric perturbation.

{\bf LHS 1749AB:} A close visual binary discovered by
\citet{2003AJ....125..332J} with a separation of 2\farcs9 at a
position angle of 140 deg (see Table \ref{tab:multiples}).  The
parallax in Table \ref{tab:astrometry} was calculated for the A
component only, and that distance (21.7$\pm$0.7 pc) is consistent with
22.0$\pm$2.5 pc reported by the CTIOPI 1.5m program
\citep{2006AJ....132.1234C}.  A weighted mean parallax for the system
is given in Table \ref{tab:weightedmeans}.

\begin{figure}
\center
\includegraphics[scale=.6]{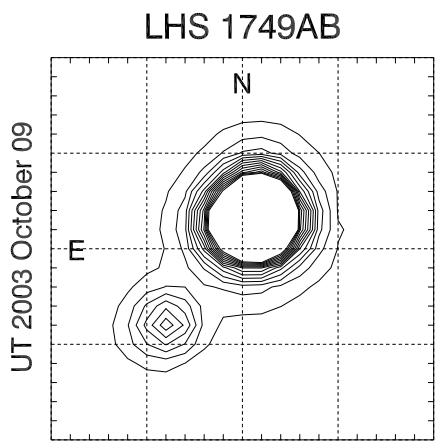}
\caption{Contour plot of LHS 1749AB in the $V$ filter on 2003 October
  09.  The B component is obvious in the image but difficult to
  separate cleanly on most frames.  The four years of available data
  suggest slight orbital motion.  Grid markings are 5 pixels
  (2.05$\arcsec$). \label{fig:lhs1749ab}}
\end{figure}

The B component is $\sim$2.8 mag fainter in $V$ than A, as shown in
Figure \ref{fig:lhs1749ab}. LHS 1749B is separable from A on 15
parallax frames; the resulting relative parallax result is
43.17$\pm$4.33 mas (23.2$\pm$2.3 pc) which is of poor quality but
consistent with other measurements.  Evidence for orbital motion is
shown in Table \ref{tab:multiples}.

{\bf LHS 1807:} A 1.5-$\sigma$ underluminous system (14.1$\pm$0.3 pc
trig/19.1$\pm$3.0 pc phot dist), but still evidently a main sequence
star (see Figure \ref{fig:HRdiagram}).

{\bf LHS 1955AB:} A close visual binary listed in
\citet{1979lcse.book.....L} with a 0\farcs8 separation at an angle of
290 deg.  Our astrometric reduction uses relaxed ellipticity
constraints (60\%, not 20\%) to keep frames where B extends the PSF of
A. The B component is within the photometric aperture and causes the
3.1-$\sigma$ overluminosity (13.5$\pm$0.2 pc trig/8.6$\pm$1.4 pc phot
dist) and the elevation above the main sequence seen in Figure
\ref{fig:HRdiagram}.  We detect no astrometric perturbation of A
despite the motion of the B component.

\begin{figure}
\center
\includegraphics[scale=.6]{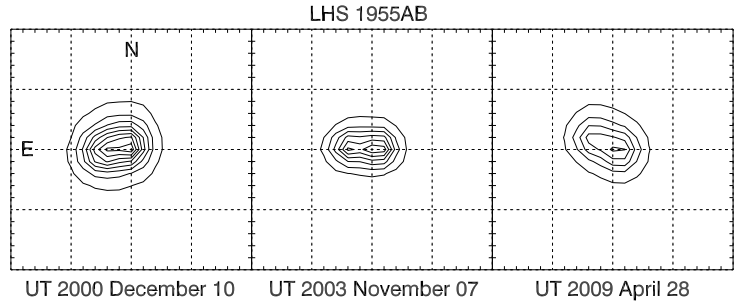}
\caption{Contour plots of LHS 1955AB for three nights in the $R$
  filter.  LHS 1955B is occasionally visible as a saddle point or even
  a peak (middle frame).  The motion seen here suggests an $\sim$80 yr
  orbit.  Grid markings are 5 pixels (2.05$\arcsec$). 
  \label{fig:lhs1955ab}}
\end{figure}

LHS 1955B is $\sim$0.5 magnitudes fainter in $R$ than A, and separable
from A on only seven frames over five nights using special SExtractor
settings.  Using those frames, we can obtain a relative parallax for
B: 73.65$\pm$19.58 mas (13.58$\pm$3.61 pc), consistent with the
relative parallax of A in Table \ref{tab:astrometry}, 72.76$\pm$1.09
mas (13.74$\pm$0.21 pc). Considerable orbital motion can be seen in
Figure \ref{fig:lhs1955ab} and Table \ref{tab:multiples} suggesting
P$\sim$80 yr. All results for B are questionable due to severe PSF
contamination.

{\bf LHS 2010:} This system has a 3.0-$\sigma$ distance mismatch
(13.7$\pm$0.3 pc trig/8.9$\pm$1.4 pc phot dist) and is elevated above
the main sequence in Figure \ref{fig:HRdiagram}.  We see no
astrometric perturbation, but it may be a multiple to which CTIOPI is
not sensitive.

{\bf LHS 2021:} Contains the lowest luminosity star in our sample: $V
=$ 19.17, spectral type M6.0V (despite unusually red colors), which
can be seen in the lower right of Figure \ref{fig:HRdiagram}. This
paper's distance (15.7$\pm$0.3 pc) is consistent with the 16.7$\pm$1.3
pc distance reported by the CTIOPI 1.5m program in
\citep{2006AJ....132.1234C}. A weighted mean system parallax is given
in Table \ref{tab:weightedmeans}.  This system is plotted as an
example of single-star astrometric residuals in Figure
\ref{fig:lhs2021}.

{\bf LHS 2071AB:} A new astrometric binary with P$>$9 years and a 21
mas photocentric semi-major axis (see Figure \ref{fig:lhs2071ab}). The
unseen companion is causing a 2.2-$\sigma$ overluminosity
(15.0$\pm$0.3 pc trig/10.8$\pm$1.7 pc phot dist, see Figure
\ref{fig:distances}) and a noticeable elevation above the main
sequence in Figure \ref{fig:HRdiagram}. The system has been resolved
with adaptive optics on Gemini North; further results will follow in a
later paper.  An orbit consistent with our current dataset is given in
Table \ref{tab:orbits}, the orbital motion was removed from the data
before fitting the final parallax.

{\bf LHS 5156:} The final parallax is entirely based on new $V$ filter
data due to insufficient old $V$ coverage.  Our reduction does not
show the characteristic new $V$ wobble ($\S$\ref{sec:astrometryobs})
in the residuals, but it may be inaccurate by more than 1-$\sigma$.

{\bf GJ 438:} The hottest and most luminous star in the sample, as can
be seen in Figure \ref{fig:HRdiagram}. The YPC distance (8.4$\pm$1.1
pc) is inconsistent with our distance, 10.9$\pm$0.3 pc.  This system
is not in the RECONS 10 parsec sample.  A weighted mean parallax to
this system is given in Table \ref{tab:weightedmeans}.

{\bf LHS 2520:} The third-most overluminous system in this sample; it
has a 3.2-$\sigma$ distance mismatch (12.8$\pm$0.4 pc trig/7.6$\pm$1.2
pc phot dist, see Figure \ref{fig:distances}) and is elevated above
the main sequence as shown in Figure \ref{fig:HRdiagram}. We detect no
astrometric perturbation; it may be a multiple to which CTIOPI is not
sensitive.

{\bf LHS 2567/2568:} A visual binary with a separation of 8\farcs0 at
a position angle of 61 deg (see Table \ref{tab:multiples}).  The A
component (LHS 2567) has a 2.7-$\sigma$ distance mismatch
(21.4$\pm$0.8 pc trig/13.6$\pm$2.1 pc phot dist, see Figure
\ref{fig:distances}) while the B component (LHS 2568) distance matches
to 0.4-$\sigma$ (20.6$\pm$0.8 pc trig/19.0$\pm$2.9 pc phot dist).  LHS
2567 shows no astrometric perturbation, but given that it should be
the same age and metallicity as LHS 2568, it is potentially an
unresolved binary much like LHS 4009AB, below.  The proper motions of
A and B are discrepant by 11.1-$\sigma$ due to orbital motion
presented in Table \ref{tab:multiples}. A weighted system parallax is
given in Table \ref{tab:weightedmeans}.

{\bf LHS 3001/3002:} A visual binary with a separation of 12\farcs7 at
a position angle of 43.9 deg (see Table \ref{tab:multiples}).  The B
component (LHS 3002) is the second-coolest star in this sample, as
shown in Figure \ref{fig:HRdiagram}.  The proper motions of A and B
are discrepant by 4.8-$\sigma$ due to orbital motion presented in
Table \ref{tab:multiples}. A weighted system parallax is given in
Table \ref{tab:weightedmeans}.

{\bf LHS 3080:} The second-most overluminous system in this sample. It
has a 3.2-$\sigma$ distance mismatch system (28.2$\pm$1.5 pc
trig/15.7$\pm$2.4 pc phot dist, see Figure \ref{fig:distances}) and is
elevated above the main sequence as shown in Figure
\ref{fig:HRdiagram}.  We detect no astrometric perturbation; it may be
a multiple to which CTIOPI is not sensitive.

{\bf LHS 3197:} We have used an average correction to absolute
parallax (1.50$\pm$0.50 mas) for this system because the calculated
correction (3.44 mas) was abnormally large.  This is likely due to
artificial reddening of the reference stars, caused by the molecular
cloud \object{LDN 1781} \citep{1962ApJS....7....1L}, which (if
circular) has radius 37$\arcmin$ and a center only 22$\arcmin$ away at
a position angle of 43 deg.

{\bf GJ 633:} The published YPC distance (9.6$\pm$1.3 pc) is
inconsistent with our distance, 16.8$\pm$0.3 pc, which supersedes the
22.5 pc$\pm$0.9 pc distance published by us in
\citet{2006AJ....132.2360H}.  Improvements in centroiding discussed in
$\S$\ref{sec:pi} now reliably distinguish GJ 633 from a point source
7\arcsec~away that contaminated the previous result.  The system is
still not in the RECONS 10 parsec sample.  A weighted mean system
parallax (this new result and YPC) is given in Table
\ref{tab:weightedmeans}.

{\bf WT 562:} Unrelated to the system \object{SCR 1826-6542}
\citep{2007AJ....133.2898F}, 5\farcm8 away.  WT 562 has $\mu =$
0\farcs611 yr$^{-1}$ at 180.9 deg while SCR 1826-6542 has $\mu =$
0\farcs311 yr$^{-1}$ at 178.9 deg.  Early results also suggest SCR
1826-6542 is several parsecs closer.

{\bf LHS 3739/3738AB:} A heirarchical triple system consisting of a
new astrometric binary, LHS 3738AB, which is itself the B component of
a known visual binary with LHS 3739.  The system is the most
underluminous in our sample. Using identical reference fields and
frames (see Figure \ref{fig:lhs3739}), LHS 3739 has no signs of
perturbation and is 1.6-$\sigma$ \emph{under}luminous (19.6$\pm$0.4 pc
trig/27.6$\pm$4.5 pc phot dist, see Figure \ref{fig:distances}) while
the light of the components of LHS 3738AB combine to give only a
0.3-$\sigma$ (19.7$\pm$0.4 pc trig/18.5$\pm$3.0 pc phot dist)
difference from expectations.  Even so, LHS 3739 (and therefore LHS
3738 A and B) seems to be main-sequence in Figure \ref{fig:HRdiagram}.

The LHS 3739/LHS 3738AB visual binary has a separation of 113\farcs1
at a position angle of 95.7 deg, and has proper motions discrepant by
2.3-$\sigma$. This orbital motion is detected and presented in Table
\ref{tab:multiples}.  A weighted system parallax is given in Table
\ref{tab:weightedmeans}.

The LHS 3738AB new astrometric binary has a 5.8 year period and a 27
mas photocentric semi-major axis (see Figure \ref{fig:lhs3739}), and
has been resolved by Gemini North. A preliminary orbit is given in
Table \ref{tab:orbits} and was removed from the data before fitting
the final parallax.  Further results will be published in a later
paper.

{\bf LHS 4009AB:} We do not confirm the companion from
\citet{2006A&A...460L..19M} resolved with adaptive optics on 14
October 2005 with a separation of 0\farcs07 at a position angle of 250
deg, and $\Delta K = 0.15$ in what \citet{2006A&A...460L..19M} claim
is a three year orbit.  The system has a 1.9-$\sigma$ distance
mismatch (12.5$\pm$0.2 pc trig/9.2$\pm$1.5 pc phot dist) and is
elevated above the main sequence, (see Figure \ref{fig:HRdiagram}) but
we detect no astrometric perturbation (see Figure
\ref{fig:lhs4009ab}), probably because the system components are
nearly equal luminosity in $R$.

{\bf LHS 4016:} The system has a 2.0-$\sigma$ distance mismatch
(24.2$\pm$0.9 pc trig/17.2$\pm$2.7 pc phot dist, see Figure
\ref{fig:distances}) and is elevated above the main sequence as shown
in Figure \ref{fig:HRdiagram}. There are possible signs of an
astrometric perturbation, but a gap from 2005 to 2009 when the old $V$
filter was not used prevents any definite determination.

\section{Discussion}
\label{sec:results}

As shown in Figure \ref{fig:precision}, our parallax errors compare
favorably to the errors from other ground-based parallax efforts, as
summarized in YPC.  Our increased accuracy can be attributed to our
use of CCD images for our astrometry, while most of the YPC parallaxes
were measured from photographic plates.

\begin{figure}
\center
\includegraphics[angle=90,scale=.4]{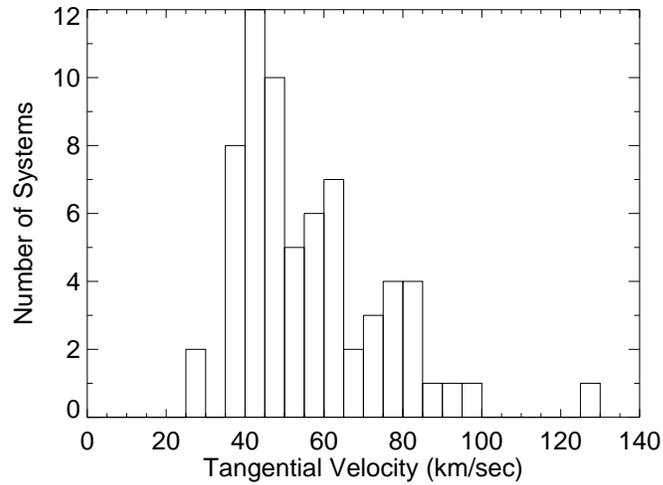}
\caption{Tangential velocity distribution for our systems.  The
  fastest moving star is also our most distant, LTT 5066. 
  \label{fig:vtandiagram}}
\end{figure}

In Figure \ref{fig:vtandiagram} we plot the distribution of tangential
velocities listed in Column 15 of Table \ref{tab:astrometry}.  Most of
the stars have $v_{tan} =$ 25--100 km sec$^{-1}$, as expected for disk
red dwarfs \citep{1981gask.book.....M}.  The single star with $v_{tan}
=$ 126 km sec$^{-1}$ is LTT 5066, which at 46 pc is the furthest star
discussed in this paper; by photometry and spectroscopy it is a dwarf,
not a subdwarf.  Our sample is kinematically biased, requiring stars
to have 0\farcs5 $\leq \mu \leq$ 1\farcs0 yr$^{-1}$. As such, the
nearest star, LHS 5156, must have a tangential velocity between 25 and
50 km sec$^{-1}$, while our farthest star, LTT 5066, could not be in
our proper motion regime if it were moving any slower than 110 km
sec$^{-1}$.

The 56 systems within 25 pc described here constitute 2.7\% of all
systems now confirmed by parallax to be in the 25 pc sample (5.7\% of
systems in the southern hemisphere), according to the statistics from
the NStars Database \citep{2002AJ....123.2002H}, using 2011 previously
known systems as a baseline.  Including parallax results from the
entire CTIOPI program, we have added 155 new systems (a 7.7\%
increase) to the all-sky 25 pc sample.  Of those 25 pc systems, 142
are in the southern hemisphere, a net increase of 16.6\% in the south
alone. We are currently observing roughly a hundred additional systems
that may prove to be within 25 pc, and continue to add more as
observing time and stamina permit.  Nonetheless, we do not anticipate
completing the census of all systems in the solar neighborhood through
CTIOPI alone, which means that ground-based sky survey efforts such as
Pan-STARRS (The Panoramic Survey Telescope \& Rapid Response System),
SkyMapper, LSST (the Large Synoptic Survey Telescope), and space-based
missions like Gaia, will undoubtedly reveal many more of the Sun's
neighbors.

\section{Acknowledgements}
\acknowledgements

The RECONS effort is supported primarily by the National Science
Foundation through grants AST 05-07711 and AST 09-08402, as well as
through NASA's {\it Space Interferometry Mission}.  Observations were
initially made possible by NOAO's Survey Program and have continued
via the SMARTS Consortium.  This research has made use of results from
the NStars Project, NASA's Astrophysics Data System Bibliographic
Services, the SIMBAD and VizieR databases operated at CDS, Strasbourg,
France, the SuperCOSMOS Sky Survey, and the 2MASS database.

The authors also wish to thank Mr. Sergio Dieterich for the Gemini
observations and reductions; Dr. Brian Mason for supplying the
orbit-fitting code; and the staff of the Cerro Tololo Inter-american
Observatory, particularly Edgardo Cosgrove, Arturo Gomez, Alberto
Miranda, and Joselino Vasquez for their help over the years.  The
authors also wish to thank Dr. Hugh Harris and Dr. Jennifer Bartlett
for their constructive comments.



\clearpage

\thispagestyle{empty}

\voffset040pt{
\begin{deluxetable}{lcccccccccrrcccl}
\rotate
\setlength{\tabcolsep}{0.02in}
\tablewidth{0pt}
\tabletypesize{\scriptsize}
\tablecaption{Astrometric Results\label{tab:astrometry}}
\tablehead{\colhead{}              &
	   \colhead{R.A.}          &
 	   \colhead{Decl.}         &
	   \colhead{}              &
	   \colhead{}              &
	   \colhead{}              &
	   \colhead{}              &
	   \colhead{}              &
	   \colhead{}              &
	   \colhead{$\pi$(rel)}    &
	   \colhead{$\pi$(corr)}   &
	   \colhead{$\pi$(abs)}    &
	   \colhead{$\mu$}         &
	   \colhead{P.A.}          &
	   \colhead{V$_{tan}$}     &
	   \colhead{}              \\

           \colhead{Name}          &
	   \multicolumn{2}{c}{(J2000.0)\tablenotemark{a}} &
	   \colhead{Filter}        &
	   \colhead{N$_{sea}$\tablenotemark{b}}&
	   \colhead{N$_{frm}$}     &
	   \colhead{Coverage\tablenotemark{b}}&
	   \colhead{Years\tablenotemark{b}}&
 	   \colhead{N$_{ref}$}     &
	   \colhead{(mas)}         &
	   \colhead{(mas)}         &
	   \colhead{(mas)}         &
	   \colhead{(mas yr$^{-1}$)}&
	   \colhead{(deg)}         &
	   \colhead{(km s$^{-1}$)} &
	   \colhead{Notes}         \\

           \colhead{(1)}           &
           \colhead{(2)}           &
           \colhead{(3)}           &
           \colhead{(4)}           &
           \colhead{(5)}           &
           \colhead{(6)}           &
           \colhead{(7)}           &
           \colhead{(8)}           &
           \colhead{(9)}           &
           \colhead{(10)}          &
           \colhead{(11)}          &
           \colhead{(12)}          &
           \colhead{(13)}          &
           \colhead{(14)}          &
           \colhead{(15)}          &
           \colhead{(16)}          }

\startdata
\hline
\object{LHS 1050}  & 00 15 49.25 & +13 33 22.3 & $V$ &  7s  &  60 & 1999.71--2009.75 & 10.03 &  7 & 84.46$\pm$2.56 & 1.39$\pm$0.22 & 85.85$\pm$2.57 & 699.2$\pm$0.5 & 061.7$\pm$0.08 &  38.6 & \tablenotemark{c}\\
\object{LHS 1351}  & 02 11 18.06 & -63 13 41.0 & $V$ &  3c+ &  68 & 2000.58--2004.97 &  4.40 &  5 & 70.62$\pm$1.64 & 0.91$\pm$0.09 & 71.53$\pm$1.64 & 764.6$\pm$1.3 & 243.2$\pm$0.18 &  50.7 & \\
\object{LHS 1358}  & 02 12 54.63 & +00 00 16.8 & $R$ &  5s  &  58 & 1999.71--2003.86 &  4.15 &  5 & 64.09$\pm$2.07 & 1.18$\pm$0.13 & 65.27$\pm$2.07 & 558.3$\pm$1.3 & 086.1$\pm$0.20 &  40.5 & \\
\object{LHS 1491}  & 03 04 04.49 & -20 22 43.0 & $V$ &  5s  &  70 & 1999.71--2005.00 &  5.29 &  9 & 66.32$\pm$1.24 & 0.96$\pm$0.16 & 67.28$\pm$1.25 & 685.4$\pm$0.7 & 135.2$\pm$0.12 &  48.3 & \\
\object{LHS 1582AB}& 03 43 22.08 & -09 33 50.9 & $R$ &  9s  &  70 & 2000.87--2009.63 &  8.76 &  6 & 45.50$\pm$1.43 & 1.99$\pm$0.28 & 47.49$\pm$1.46 & 506.2$\pm$0.5 & 052.3$\pm$0.10 &  50.5 & \tablenotemark{c}\tablenotemark{d}\\
\object{LHS 1630AB}& 04 07 20.50 & -24 29 13.7 & $V$ & 10s  & 137 & 1999.71--2009.03 &  9.32 &  5 & 54.44$\pm$1.04 & 1.74$\pm$0.20 & 56.18$\pm$1.06 & 683.1$\pm$0.4 & 163.7$\pm$0.05 &  57.6 & \tablenotemark{c}\tablenotemark{e}\\
\object{LHS 1748}  & 05 15 46.72 & -31 17 45.3 & $V$ &  5c  &  58 & 2000.88--2005.14 &  4.26 &  9 & 41.76$\pm$1.40 & 1.42$\pm$0.05 & 43.18$\pm$1.40 & 551.6$\pm$1.1 & 062.6$\pm$0.22 &  60.6 & \\
\object{LHS 1749A} & 05 16 00.39 & -72 14 12.6 & $V$ &  5s  &  91 & 2000.88--2005.05 &  4.17 &  6 & 44.93$\pm$1.46 & 1.21$\pm$0.08 & 46.14$\pm$1.46 & 814.2$\pm$1.1 & 356.4$\pm$0.12 &  83.6 & \tablenotemark{c}\tablenotemark{g}\\
\object{LHS 1767}  & 05 31 04.33 & -30 11 44.8 & $V$ &  5c  &  57 & 2003.96--2007.99 &  4.03 & 10 & 64.53$\pm$1.51 & 0.73$\pm$0.05 & 65.26$\pm$1.51 & 580.1$\pm$1.1 & 143.8$\pm$0.22 &  42.1 & \tablenotemark{e}\\
\object{WT 178}    & 05 37 39.77 & -61 54 43.8 & $R$ &  5s+ &  68 & 1999.91--2003.77 &  3.86 & 10 & 60.53$\pm$0.89 & 1.49$\pm$0.09 & 62.02$\pm$0.89 & 502.9$\pm$0.9 & 015.8$\pm$0.19 &  38.4 & \\
\object{APMPM J0544-4108}&05 43 46.56&-41 08 08.4&$V$&  5c  &  76 & 2000.14--2005.05 &  4.91 &  8 & 47.63$\pm$0.78 & 0.78$\pm$0.08 & 48.41$\pm$0.78 & 601.4$\pm$0.4 & 165.6$\pm$0.07 &  58.9 & \\
\object{LHS 1807}  & 06 02 22.62 & -20 19 44.2 & $R$ &  4c+ &  66 & 2000.88--2007.83 &  6.95 &  6 & 70.02$\pm$1.57 & 0.98$\pm$0.14 & 71.00$\pm$1.58 & 558.3$\pm$1.3 & 355.7$\pm$0.19 &  37.3 & \tablenotemark{c}\\
\object{GJ 1088}   & 06 10 52.89 & -43 24 17.8 & $V$ &  5s  &  51 & 2000.88--2005.06 &  4.18 &  6 & 85.44$\pm$1.27 & 1.59$\pm$0.19 & 87.03$\pm$1.28 & 745.5$\pm$0.8 & 010.6$\pm$0.11 &  40.6 & \\
\object{LHS 1932}  & 07 36 12.03 & -51 55 21.3 & $V$ &  3c  &  93 & 2000.88--2003.14 &  2.26 &  9 & 60.91$\pm$0.98 & 1.01$\pm$0.09 & 61.92$\pm$0.98 & 595.1$\pm$1.2 & 042.8$\pm$0.22 &  45.6 & \\
\object{LHS 1955A} & 07 54 54.80 & -29 20 56.4 & $R$ &  5c  &  83 & 2000.94--2009.32 &  8.38 & 10 & 72.76$\pm$1.09 & 1.60$\pm$0.29 & 74.36$\pm$1.13 & 596.6$\pm$0.5 & 146.8$\pm$0.09 &  38.0 & \tablenotemark{c}\tablenotemark{g}\\
\object{LHS 2010}  & 08 27 11.83 & -44 59 21.1 & $V$ &  4c  &  85 & 2001.14--2004.18 &  3.03 &  9 & 71.33$\pm$1.27 & 1.47$\pm$0.30 & 72.80$\pm$1.30 & 537.0$\pm$1.2 & 343.1$\pm$0.22 &  35.0 & \tablenotemark{c}\\
\object{LHS 2021}  & 08 30 32.57 & +09 47 15.5 & $I$ &  6s+ &  54 & 2003.94--2009.24 &  5.30 &  8 & 62.06$\pm$1.14 & 1.61$\pm$0.18 & 63.67$\pm$1.15 & 667.2$\pm$0.6 & 226.6$\pm$0.10 &  49.7 & \tablenotemark{c}\\
\object{LHS 2071AB}& 08 55 20.25 & -23 52 15.0 & $R$ &  7s  &  69 & 2000.07--2009.30 &  9.23 & 11 & 66.16$\pm$1.25 & 0.65$\pm$0.08 & 66.81$\pm$1.25 & 591.5$\pm$0.3 & 276.5$\pm$0.06 &  42.0 & \tablenotemark{c}\tablenotemark{d}\\
\object{LHS 2106}  & 09 07 02.75 & -22 08 50.1 & $R$ &  5s  &  56 & 2000.06--2006.04 &  5.97 &  7 & 65.12$\pm$1.16 & 1.11$\pm$0.06 & 66.23$\pm$1.16 & 505.7$\pm$0.4 & 215.4$\pm$0.08 &  36.2 & \\
\object{LHS 2122}  & 09 16 25.99 & -62 04 16.0 & $R$ &  6s+ &  64 & 2001.15--2009.04 &  7.89 &  9 & 56.42$\pm$2.57 & 2.17$\pm$0.15 & 58.59$\pm$2.57 & 933.8$\pm$0.9 & 313.3$\pm$0.10 &  75.5 & \\
\object{LHS 5156}  & 09 42 49.60 & -63 37 56.1 & $V$ &  4s+ &  60 & 2005.97--2009.32 &  3.35 &  9 & 94.42$\pm$1.17 & 0.73$\pm$0.07 & 95.15$\pm$1.17 & 516.6$\pm$0.9 & 084.5$\pm$0.15 &  25.7 & \tablenotemark{c}\tablenotemark{f}\\
\object{WT 244}    & 09 44 23.73 & -73 58 38.3 & $I$ &  7s+ &  64 & 1999.92--2008.00 &  8.08 & 12 & 42.05$\pm$1.48 & 1.25$\pm$0.23 & 43.30$\pm$1.50 & 517.6$\pm$0.5 & 258.1$\pm$0.08 &  56.7 & \\
\object{LHS 2328}  & 10 55 34.47 & -09 21 25.9 & $R$ &  8s  &  66 & 2001.15--2009.25 &  8.10 &  9 & 53.03$\pm$1.47 & 0.81$\pm$0.08 & 53.84$\pm$1.47 & 516.2$\pm$0.5 & 330.9$\pm$0.10 &  45.4 & \\
\object{LHS 2335}  & 10 58 35.10 & -31 08 38.4 & $V$ &  3c  &  56 & 2001.14--2003.14 &  1.99 &  8 & 49.58$\pm$1.54 & 0.97$\pm$0.20 & 50.55$\pm$1.55 & 571.7$\pm$3.0 & 260.9$\pm$0.49 &  53.6 & \\
\object{LHS 2401}  & 11 23 57.31 & -18 21 48.6 & $V$ &  5s  &  70 & 2001.15--2005.14 &  3.99 &  5 & 51.74$\pm$2.50 & 2.73$\pm$0.17 & 54.47$\pm$2.51 & 576.4$\pm$1.4 & 266.6$\pm$0.22 &  50.2 & \\
\object{GJ 1147}   & 11 38 24.95 & -41 22 32.5 & $R$ &  5s+ &  64 & 2001.15--2009.04 &  7.89 & 10 & 65.13$\pm$1.08 & 0.95$\pm$0.09 & 66.08$\pm$1.08 & 942.4$\pm$0.7 & 274.0$\pm$0.06 &  67.6 & \\
\object{GJ 438}    & 11 43 19.82 & -51 50 25.9 & $V$ &  6s+ &  92 & 2000.06--2009.32 &  9.26 &  6 & 89.13$\pm$2.03 & 2.57$\pm$0.30 & 91.70$\pm$2.05 & 857.0$\pm$0.9 & 129.9$\pm$0.12 &  44.3 & \tablenotemark{c}\tablenotemark{e}\\
\object{LHS 2520}  & 12 10 05.59 & -15 04 16.9 & $V$ &  4s  &  56 & 2000.07--2004.43 &  4.37 &  7 & 77.11$\pm$2.40 & 0.82$\pm$0.17 & 77.93$\pm$2.41 & 718.6$\pm$1.6 & 183.5$\pm$0.18 &  43.7 & \\
\object{GJ 1157}   & 12 23 01.43 & -46 37 08.4 & $V$ &  5s  &  61 & 2001.14--2005.14 &  3.99 &  7 & 61.00$\pm$0.61 & 1.42$\pm$0.16 & 62.42$\pm$0.63 & 819.5$\pm$0.5 & 245.0$\pm$0.06 &  62.2 & \\
\object{LHS 2567}  & 12 29 54.19 & -05 27 24.4 & $R$ &  7s  &  58 & 2000.07--2009.03 &  8.96 &  7 & 45.54$\pm$1.83 & 1.26$\pm$0.07 & 46.80$\pm$1.83 & 611.1$\pm$0.6 & 241.0$\pm$0.11 &  61.9 & \tablenotemark{c}A \\
\object{LHS 2568}  & 12 29 54.66 & -05 27 20.6 & $R$ &  7s  &  58 & 2000.07--2009.03 &  8.96 &  7 & 47.29$\pm$1.81 & 1.26$\pm$0.07 & 48.55$\pm$1.81 & 597.8$\pm$0.6 & 241.5$\pm$0.11 &  58.4 & \tablenotemark{c}B \\
\object{LHS 2718}  & 13 20 03.86 & -35 24 44.1 & $V$ &  5s  &  62 & 2001.15--2005.14 &  3.99 & 11 & 72.05$\pm$0.78 & 0.99$\pm$0.11 & 73.04$\pm$0.79 & 961.6$\pm$0.6 & 241.1$\pm$0.07 &  62.4 & \\
\object{LHS 2729}  & 13 23 38.02 & -25 54 45.1 & $R$ &  5s  &  56 & 2001.15--2005.09 &  3.94 & 12 & 70.32$\pm$1.52 & 1.16$\pm$0.14 & 71.48$\pm$1.53 & 633.9$\pm$1.3 & 249.8$\pm$0.21 &  42.0 & \\
\object{LHS 2836}  & 13 59 10.45 & -19 50 03.4 & $V$ &  3c+ & 108 & 2000.14--2004.18 &  4.04 &  8 & 91.22$\pm$0.86 & 1.64$\pm$0.23 & 92.86$\pm$0.89 & 573.4$\pm$1.0 & 252.0$\pm$0.17 &  29.3 & \\
\object{LHS 2899}  & 14 21 15.12 & -01 07 19.7 & $V$ &  5s  &  45 & 2000.14--2005.14 &  4.99 &  8 & 74.00$\pm$2.15 & 0.66$\pm$0.05 & 74.66$\pm$2.15 & 643.5$\pm$1.4 & 164.4$\pm$0.21 &  40.9 & \\
\object{LHS 3001}  & 14 56 27.16 & +17 57 00.0 & $I$ &  5s  &  78 & 2000.58--2009.25 &  8.67 &  9 & 56.36$\pm$1.34 & 0.81$\pm$0.07 & 57.17$\pm$1.34 & 982.4$\pm$0.8 & 301.2$\pm$0.09 &  81.5 & \tablenotemark{c}A \\
\object{LHS 3002}  & 14 56 27.79 & +17 55 08.9 & $I$ &  5s  &  78 & 2000.58--2009.25 &  8.67 &  9 & 54.83$\pm$1.35 & 0.81$\pm$0.07 & 55.64$\pm$1.35 & 987.8$\pm$0.8 & 301.5$\pm$0.09 &  84.2 & \tablenotemark{c}B \\
\object{LHS 3167}  & 16 13 05.93 & -70 09 08.0 & $R$ &  6s  &  85 & 2000.57--2009.32 &  8.75 & 10 & 58.22$\pm$0.93 & 2.03$\pm$0.21 & 60.25$\pm$0.95 & 607.1$\pm$0.5 & 202.1$\pm$0.08 &  47.8 & \\
\object{LHS 3169}  & 16 14 21.93 & -28 30 36.7 & $V$ &  5s  &  53 & 2000.58--2004.45 &  3.87 & 10 & 52.03$\pm$1.45 & 1.40$\pm$0.17 & 53.43$\pm$1.46 & 523.9$\pm$1.0 & 230.8$\pm$0.22 &  46.5 & \\
\object{LHS 3197}  & 16 26 48.12 & -17 23 34.3 & $R$ &  4c+ &  51 & 2000.23--2006.37 &  6.14 &  9 & 53.53$\pm$1.27 & 1.50$\pm$0.50\tablenotemark{h}&55.03$\pm$1.36&522.2$\pm$1.0&219.0$\pm$0.22&45.0&\tablenotemark{c}\\
\object{LHS 3218}  & 16 35 24.64 & -27 18 54.7 & $R$ &  6s  &  77 & 2000.23--2009.32 &  9.09 &  8 & 51.31$\pm$0.74 & 2.57$\pm$0.23 & 53.88$\pm$0.77 & 889.0$\pm$0.3 & 180.7$\pm$0.03 &  78.2 & \\
\object{GJ 633}    & 16 40 45.26 & -45 59 59.3 & $V$ &  7s+ & 100 & 1999.64--2007.44 &  7.80 & 10 & 56.88$\pm$1.05 & 2.59$\pm$0.56 & 59.47$\pm$1.19 & 527.2$\pm$0.4 & 137.4$\pm$0.09 &  42.0 & \tablenotemark{c}\tablenotemark{e}\\
\object{LHS 3295}  & 17 29 27.34 & -80 08 57.4 & $V$ &  5s  &  68 & 2000.57--2004.25 &  3.68 &  8 & 78.20$\pm$1.79 & 1.75$\pm$0.33 & 79.95$\pm$1.82 & 701.7$\pm$1.5 & 312.8$\pm$0.25 &  41.6 & \\
\object{WT 562}    & 18 26 19.80 & -65 47 41.1 & $I$ &  4c+ &  80 & 2000.58--2005.64 &  5.06 &  7 & 57.53$\pm$0.90 & 0.90$\pm$0.07 & 58.43$\pm$0.90 & 610.8$\pm$1.1 & 180.9$\pm$0.15 &  49.6 & \tablenotemark{c}\\
\object{LHS 3413}  & 18 49 51.21 & -57 26 48.6 & $R$ &  5s  &  69 & 2000.57--2009.32 &  8.75 &  9 & 81.13$\pm$2.01 & 1.08$\pm$0.08 & 82.21$\pm$2.01 & 675.7$\pm$0.8 & 254.9$\pm$0.12 &  39.0 & \\
\object{LHS 3443}  & 19 13 07.96 & -39 01 53.8 & $V$ &  5s  &  69 & 2000.58--2009.75 &  9.17 &  8 & 47.44$\pm$1.14 & 1.13$\pm$0.10 & 48.57$\pm$1.14 & 509.2$\pm$0.5 & 118.3$\pm$0.10 &  49.7 & \\
\object{LHS 3583}  & 20 46 37.08 & -81 43 13.7 & $V$ &  8s  &  68 & 2000.58--2009.32 &  8.75 &  9 & 93.12$\pm$2.37 & 1.60$\pm$0.21 & 94.72$\pm$2.38 & 764.6$\pm$1.0 & 135.0$\pm$0.15 &  38.3 & \tablenotemark{e}\\
\object{APMPM J2127-3844}&21 27 04.58&-38 44 50.8&$R$&  4s  &  58 & 1999.62--2004.73 &  5.11 &  8 & 48.76$\pm$1.38 & 0.49$\pm$0.03 & 49.25$\pm$1.38 & 897.2$\pm$0.8 & 141.6$\pm$0.10 &  86.3 & \\
\object{WT 795}    & 21 36 25.30 & -44 01 00.2 & $V$ &  4c+ &  74 & 2000.41--2004.44 &  4.03 &  5 & 68.84$\pm$0.69 & 0.69$\pm$0.12 & 69.53$\pm$0.70 & 827.0$\pm$0.6 & 144.4$\pm$0.08 &  56.4 & \\
\object{LHS 3719}  & 21 49 25.91 & -63 06 51.9 & $V$ &  4s+ &  70 & 2000.58--2003.69 &  3.11 &  8 & 59.05$\pm$1.37 & 1.23$\pm$0.08 & 60.28$\pm$1.37 & 537.2$\pm$1.2 & 032.2$\pm$0.24 &  42.2 & \\
\object{LHS 3738AB}& 21 58 49.13 & -32 26 25.5 & $R$ &  9s  & 113 & 1999.64--2009.65 & 10.01 & 10 & 49.60$\pm$1.14 & 1.27$\pm$0.07 & 50.87$\pm$1.14 & 537.3$\pm$0.5 & 228.8$\pm$0.08 &  50.1 & \tablenotemark{c}BC\tablenotemark{d}\\
\object{LHS 3739}  & 21 58 50.19 & -32 28 17.8 & $R$ &  9s  & 113 & 1999.64--2009.65 & 10.01 & 10 & 49.70$\pm$1.05 & 1.27$\pm$0.07 & 50.97$\pm$1.05 & 535.5$\pm$0.3 & 229.2$\pm$0.07 &  49.8 & \tablenotemark{c}A \\
\object{WT 870}    & 22 06 40.68 & -44 58 07.4 & $R$ &  6s  &  70 & 2000.41--2005.90 &  5.48 &  7 & 55.41$\pm$1.13 & 1.10$\pm$0.05 & 56.51$\pm$1.13 & 736.3$\pm$0.7 & 219.6$\pm$0.11 &  61.8 & \\
\object{LHS 3909}  & 23 12 11.30 & -14 06 11.9 & $R$ &  3c+ &  55 & 2000.58--2007.55 &  6.97 &  6 & 53.41$\pm$2.00 & 1.49$\pm$0.05 & 54.90$\pm$2.00 & 716.3$\pm$1.7 & 196.0$\pm$0.24 &  61.8 & \\
\object{LHS 3925}  & 23 17 50.33 & -48 18 47.2 & $R$ &  4c  &  62 & 2000.58--2005.80 &  5.23 & 10 & 45.38$\pm$1.17 & 1.47$\pm$0.08 & 46.85$\pm$1.17 & 755.5$\pm$0.6 & 157.5$\pm$0.08 &  76.4 & \\
\object{LHS 4009AB}& 23 45 31.26 & -16 10 20.1 & $R$ &  6s  &  83 & 1999.62--2009.78 & 10.15 &  7 & 79.38$\pm$1.37 & 0.59$\pm$0.03 & 79.97$\pm$1.37 & 690.7$\pm$0.4 & 216.9$\pm$0.07 &  40.9 & \tablenotemark{c}\\
\object{LHS 4016}  & 23 48 36.06 & -27 39 38.9 & $V$ &  6s  &  68 & 2000.87--2009.75 &  8.87 &  6 & 40.75$\pm$1.54 & 0.50$\pm$0.19 & 41.25$\pm$1.55 & 595.3$\pm$0.4 & 246.2$\pm$0.08 &  68.5 & \tablenotemark{c}\\
\object{LHS 4021}  & 23 50 31.64 & -09 33 32.6 & $V$ &  4s+ &  60 & 2000.71--2004.89 &  4.18 &  6 & 61.48$\pm$1.70 & 0.93$\pm$0.04 & 62.41$\pm$1.70 & 758.1$\pm$1.4 & 121.7$\pm$0.20 &  57.6 & \\
\object{LHS 4058}  & 23 59 51.38 & -34 06 42.5 & $V$ &  5s+ &  59 & 2000.41--2006.87 &  6.45 &  7 & 61.15$\pm$1.98 & 1.99$\pm$0.38 & 63.14$\pm$2.02 & 939.0$\pm$1.2 & 132.8$\pm$0.15 &  70.5 & \tablenotemark{e}\\
\hline
\multicolumn{16}{c}{Beyond 25 pc}\\
\hline
\object{LHS 1561}  & 03 34 39.63 & -04 50 33.3 & $V$ &  6c  &  61 & 2000.07--2009.99 &  9.93 &  9 & 33.19$\pm$1.72 & 1.01$\pm$0.11 & 34.20$\pm$1.72 & 513.0$\pm$0.7 & 126.6$\pm$0.16 &  71.1 & \tablenotemark{c}\\
\object{LHS 1656}  & 04 18 51.03 & -57 14 01.1 & $I$ &  5c+ &  51 & 2003.95--2009.74 &  5.78 &  8 & 38.16$\pm$1.94 & 1.25$\pm$0.08 & 39.41$\pm$1.94 & 814.5$\pm$1.3 & 022.1$\pm$0.16 &  98.0 & \\
\object{LTT 5066}  & 13 13 29.63 & -32 27 05.3 & $R$ &  6s+ &  69 & 2000.14--2009.32 &  9.18 &  8 & 20.61$\pm$0.92 & 0.98$\pm$0.04 & 21.59$\pm$0.92 & 575.9$\pm$0.4 & 267.6$\pm$0.05 & 126.4 & \\
\object{LHS 3080}  & 15 31 54.17 & +28 51 09.7 & $R$ &  6c  &  66 & 2000.58--2009.57 &  8.99 &  9 & 34.73$\pm$1.85 & 0.79$\pm$0.06 & 35.52$\pm$1.85 & 538.8$\pm$0.7 & 274.4$\pm$0.11 &  71.9 & \tablenotemark{c}\\
\object{LHS 3147}  & 16 02 23.57 & -25 05 57.3 & $R$ &  5s+ &  72 & 2001.21--2009.31 &  8.10 &  8 & 37.19$\pm$1.35 & 1.99$\pm$0.22 & 39.18$\pm$1.37 & 666.3$\pm$0.6 & 202.3$\pm$0.09 &  80.6 & \\
\object{GJ 762}    & 19 34 36.48 & -62 50 38.6 & $V$ &  4c  &  74 & 2000.58--2003.30 &  2.72 & 10 & 37.20$\pm$1.34 & 1.52$\pm$0.10 & 38.72$\pm$1.34 & 510.7$\pm$1.5 & 227.2$\pm$0.34 &  62.5 & \\
\object{LHS 3484}  & 19 47 04.49 & -71 05 33.1 & $R$ &  7s  &  65 & 2000.58--2009.32 &  8.75 &  8 & 37.51$\pm$1.51 & 2.14$\pm$0.15 & 39.65$\pm$1.52 & 649.8$\pm$0.6 & 172.9$\pm$0.09 &  77.7 & \\
\object{LHS 3836}  & 22 38 02.92 & -65 50 08.8 & $R$ &  5s  &  61 & 1999.62--2004.45 &  4.82 &  7 & 35.90$\pm$1.32 & 0.52$\pm$0.02 & 36.42$\pm$1.32 & 693.9$\pm$1.0 & 118.9$\pm$0.17 &  90.3 & \\         
\hline
\enddata

\vfil\eject

\tablenotetext{a}{coordinates are epoch and equinox 2000.0; each
target's coordinates were extracted from 2MASS and then transformed to
epoch 2000.0 using the proper motions and position angles listed
here.}

\tablenotetext{b}{`Coverage' and `Years' run from the first to last
data point; `Seasons' counts observing semesters where a dataset was
taken, and denotes if coverage was `c'ontinuous (more than one night
of data in all seasons) or `s'cattered.  Coverage extended by a single
frame is denoted with a + in the Seasons column.}

\tablenotetext{c}{System has notes in $\S$\ref{sec:systemnotes}.}

\tablenotetext{d}{The astrometric perturbation was removed from the
final parallax fit.}

\tablenotetext{e}{Astrometric results use new $V$ filter data.}

\tablenotetext{f}{Astrometric results use new $V$ filter data {\it only}.}

\tablenotetext{g}{Parallax measured for the A component alone.}

\tablenotetext{h}{Generic correction to absolute adopted; field is
reddened by a nebula.}

\end{deluxetable}
}

\clearpage

\thispagestyle{empty}

\voffset060pt{
\begin{deluxetable}{llcccclccrrrrccccl}
\rotate
\setlength{\tabcolsep}{0.02in}
\tablewidth{0pt}
\tabletypesize{\scriptsize}
\tablecaption{Photometric Results\label{tab:photometry}}
\tablehead{\colhead{}              &
	   \colhead{Alternate}     &
	   \colhead{}              &
 	   \colhead{}              &
	   \colhead{}              &
	   \colhead{No. of abs.}   &
	   \colhead{}              &
	   \colhead{$\sigma$}      &
	   \colhead{No. of rel.}   &
	   \colhead{No. of}        &
	   \colhead{$J$}           &
	   \colhead{$H$}           &
	   \colhead{$K$}           &
	   \colhead{spectral}      &
	   \colhead{}              &
	   \colhead{phot}          &
	   \colhead{No. of}        &
	   \colhead{}              \\

           \colhead{Name}          &
           \colhead{Name}          &
	   \colhead{$V_{J}$}       &
 	   \colhead{$R_{KC}$}      &
	   \colhead{$I_{KC}$}      &
	   \colhead{Nights}        &
	   \colhead{$\pi$ filter\tablenotemark{a}}&
	   \colhead{(mag)}         &
 	   \colhead{Nights}        &
 	   \colhead{Frames}        &
	   \colhead{(2MASS)}       &
	   \colhead{(2MASS)}       &
	   \colhead{(2MASS)}       &
	   \colhead{type}          &
	   \colhead{ref}           &
	   \colhead{dist}          &
	   \colhead{Relations}     &
	   \colhead{Notes}         \\

           \colhead{(1)}           &         
           \colhead{(2)}           &
           \colhead{(3)}           &
           \colhead{(4)}           &
           \colhead{(5)}           &
           \colhead{(6)}           &
           \colhead{(7)}           &
           \colhead{(8)}           &
           \colhead{(9)}           &
           \colhead{(10)}          &
           \colhead{(11)}          &
           \colhead{(12)}          &
           \colhead{(13)}          &
           \colhead{(14)}          &
           \colhead{(15)}          &
           \colhead{(16)}          &
           \colhead{(17)}          &
           \colhead{(18)}          }

\startdata
\hline
LHS 1050   & GJ 12     & 12.62 & 11.46 & 10.04 & 3 & $V$ & .011 & 12 &  60 &  8.62 &  8.07 &  7.81 & M3.0V & ..... & 15.17$\pm$2.37 & 12 &  \\
LHS 1351   & L 125-51  & 12.23 & 11.15 &  9.82 & 2 & $V$ & .010 & 12 &  68 &  8.54 &  7.98 &  7.73 & M2.5V & Haw96\tablenotemark{b}&18.04$\pm$2.81&12& \\
LHS 1358   & G 159-46  & 13.58 & 12.31 & 10.66 & 2 & $R$ & .015 & 11 &  58 &  9.06 &  8.52 &  8.17 & M4.0V & Rei95\tablenotemark{c}&12.54$\pm$1.94&12& \\
LHS 1491   & LP 771-77 & 12.84 & 11.65 & 10.13 & 2 & $V$ & .018 & 15 &  70 &  8.63 &  8.02 &  7.75 & M3.5V & Rei95 & 12.42$\pm$1.95 & 12 &  \\
LHS 1582AB & G 160-19  & 14.69 & 13.33 & 11.60 & 4 & $R$ & .019 & 19 &  70 &  9.80 &  9.18 &  8.85 & M4.5V & Rei95 & 13.27$\pm$2.25 & 12 &  \\
LHS 1630AB & LP 833-42 & 12.38 & 11.22 &  9.68 & 4 & $V$ & .015 & 22 & 137 &  8.24 &  7.68 &  7.44 & M3.5VJ& ..... & 11.72$\pm$1.84 & 12 & \tablenotemark{a}\\
LHS 1748   & L 521-2   & 12.08 & 11.06 &  9.83 & 2 & $V$ & .017 & 11 &  58 &  8.59 &  7.99 &  7.73 & M2.5V & Haw96 & 19.91$\pm$3.13 & 12 &  \\
LHS 1749A  & L 57-44   & 11.74 & 10.70 &  9.47 & 2 & $V$ & .028 & 16 &  91 &  8.21 &  7.62 &  7.36 & M2.0V & Haw96 & 16.44$\pm$2.58 & 12 &  \\
LHS 1767   & LP 892-51 & 13.11 & 11.93 & 10.45 & 2 & $V$ & .012 & 13 &  57 &  9.05 &  8.49 &  8.19 & M3.0V & ..... & 17.31$\pm$2.72 & 12 & \tablenotemark{a}\\
WT 178     &           & 14.81 & 13.47 & 11.77 & 3 & $R$ & .014 & 16 &  68 & 10.14 &  9.53 &  9.23 & M4.5V & Rei07\tablenotemark{d}&18.41$\pm$2.86&12&\\
APMPM J0544-4108&      & 14.12 & 12.85 & 11.25 & 2 & $V$ & .010 & 17 &  76 &  9.74 &  9.16 &  8.87 & M3.5V & ..... & 19.08$\pm$3.01 & 12 &  \\
LHS 1807   & LP 779-10 & 13.26 & 12.10 & 10.62 & 2 & $R$ & .008 & 12 &  66 &  9.22 &  8.67 &  8.37 & M3.0V & Kir95\tablenotemark{e}&19.12$\pm$2.98&12&\\
GJ 1088    & LHS 1831  & 12.28 & 11.11 &  9.61 & 2 & $V$ & .016 & 13 &  51 &  8.17 &  7.58 &  7.31 & M3.5V & ..... & 11.02$\pm$1.69 & 12 &  \\
LHS 1932   & L 240-16  & 12.48 & 11.36 &  9.92 & 2 & $V$ & .010 & 16 &  93 &  8.55 &  8.04 &  7.76 & M3.5V & Haw96 & 15.84$\pm$2.52 & 12 &  \\
LHS 1955AB & L 601-78  & 12.79 & 11.52 &  9.89 & 2 & $R$ & .011 & 15 &  83 &  8.31 &  7.69 &  7.35 & M4.0V & Rei95 &  8.56$\pm$1.35 & 12 &  \\
LHS 2010   & L 387-102 & 11.86 & 10.70 &  9.19 & 3 & $V$ & .011 & 14 &  85 &  7.75 &  7.15 &  6.87 & M3.5V & Haw96 &  8.91$\pm$1.37 & 12 &  \\
LHS 2021   & LP 485-17 & 19.21 & 16.94 & 14.66 & 3 & $I$ & .016 & 15 &  54 & 11.89 & 11.17 & 10.76 & M6.0V & ..... & 14.42$\pm$2.26 & 12 &  \\
LHS 2071AB & LP 844-28 & 13.88 & 12.55 & 10.82 & 3 & $R$ & .016 & 16 &  69 &  9.11 &  8.54 &  8.20 & M4.0VJ& ..... & 10.77$\pm$1.66 & 12 &  \\
LHS 2106   & LP 845-23 & 14.21 & 12.87 & 11.13 & 3 & $R$ & .014 & 12 &  56 &  9.53 &  9.00 &  8.65 & M4.5V & Rei95 & 14.47$\pm$2.42 & 12 &  \\
LHS 2122   & L 140-119 & 12.57 & 11.43 &  9.94 & 2 & $R$ & .015 & 11 &  64 &  8.47 &  7.83 &  7.55 & M3.5V & Haw96 & 11.86$\pm$2.00 & 12 &  \\
LHS 5156   & L 140-289 & 13.30 & 11.98 & 10.28 & 4 & $V$ & .010 & 14 &  60 &  8.62 &  8.10 &  7.77 & M4.5V & ..... &  9.62$\pm$1.50 & 12 & \tablenotemark{a}\\
WT 244     &           & 15.17 & 13.80 & 12.02 & 3 & $I$ & .010 & 14 &  64 & 10.23 &  9.71 &  9.38 & M4.5V & ..... & 17.13$\pm$2.66 & 12 &  \\
LHS 2328   & G 163-23  & 13.55 & 12.37 & 10.86 & 2 & $R$ & .020 & 15 &  66 &  9.42 &  8.87 &  8.61 & M3.5V & Rei95 & 20.27$\pm$3.16 & 12 &  \\
LHS 2335   & LP 905-36 & 11.93 & 10.90 &  9.63 & 2 & $V$ & .010 &  9 &  56 &  8.36 &  7.76 &  7.47 & M2.5V & Haw96 & 16.62$\pm$2.68 & 12 &  \\
LHS 2401   & L 755-53  & 13.10 & 11.97 & 10.54 & 3 & $V$ & .014 & 12 &  70 &  9.17 &  8.59 &  8.32 & M3.0V & Rei95 & 19.89$\pm$3.06 & 12 &  \\
GJ 1147    & LHS 2435  & 13.72 & 12.49 & 10.91 & 3 & $R$ & .014 & 14 &  64 &  9.44 &  8.86 &  8.54 & M3.0V & Haw96 & 17.40$\pm$2.73 & 12 &  \\
GJ 438     & LHS 2447  & 10.35 &  9.36 &  8.27 & 4 & $V$ & .008 & 14 &  92 &  7.14 &  6.58 &  6.32 & M1.0V & ..... & 12.63$\pm$1.98 & 12 & \tablenotemark{a}\\
LHS 2520   & LP 734-32 & 12.09 & 10.88 &  9.30 & 3 & $V$ & .014 & 10 &  56 &  7.77 &  7.14 &  6.86 & M3.5V & Rei95 &  7.59$\pm$1.20 & 12 &  \\
GJ 1157    & LHS 2552  & 13.59 & 12.35 & 10.71 & 2 & $V$ & .010 & 13 &  61 &  9.17 &  8.63 &  8.36 & M4.0V & Haw96 & 14.98$\pm$2.37 & 12 &  \\
LHS 2567   & G 13-44A  & 13.08 & 11.87 & 10.33 & 3 & $R$ & .015 & 12 &  58 &  8.82 &  8.27 &  7.96 & M3.5V & Rei95 & 13.55$\pm$2.09 & 12 & A \\
LHS 2568   & G 13-44B  & 14.21 & 12.96 & 11.37 & 3 & $R$ & .013 & 12 &  58 &  9.79 &  9.24 &  8.92 & M3.5V & Rei95 & 18.99$\pm$2.94 & 12 & B \\
LHS 2718   & L 473-1   & 12.84 & 11.70 & 10.24 & 3 & $V$ & .012 & 12 &  62 &  8.83 &  8.25 &  7.98 & M3.0V & Haw96 & 16.04$\pm$2.46 & 12 &  \\
LHS 2729   & L 617-35  & 12.89 & 11.68 & 10.14 & 2 & $R$ & .012 &  9 &  56 &  8.66 &  8.07 &  7.78 & M3.5V & Rei95 & 12.59$\pm$1.94 & 12 &  \\
LHS 2836   & L 763-63  & 12.88 & 11.60 &  9.90 & 3 & $V$ & .013 & 22 & 108 &  8.33 &  7.76 &  7.44 & M4.0V & ..... &  8.86$\pm$1.39 & 12 &  \\
LHS 2899   & G 124-27  & 13.12 & 11.92 & 10.39 & 3 & $V$ & .016 & 12 &  45 &  8.95 &  8.39 &  8.09 & M3.5V & Rei95 & 15.40$\pm$2.41 & 12 &  \\
LHS 3001   & LP 441-33 & 15.81 & 14.35 & 12.52 & 2 & $I$ & .013 & 17 &  78 & 10.74 & 10.15 &  9.85 & M4.5V & Rei95 & 19.74$\pm$3.12 & 12 & A \\
LHS 3002   & LP 441-34 & 18.68 & 16.65 & 14.42 & 2 & $I$ & .012 & 17 &  78 & 11.98 & 11.30 & 10.92 & M6  V & Rei95 & 17.47$\pm$2.72 & 12 & B \\
LHS 3167   & L 74-208  & 13.71 & 12.45 & 10.82 & 3 & $R$ & .014 & 17 &  85 &  9.26 &  8.74 &  8.39 & M4.0V & ..... & 14.73$\pm$2.30 & 12 &  \\
LHS 3169   & L 626-41  & 12.95 & 11.80 & 10.29 & 2 & $V$ & .011 & 10 &  53 &  8.92 &  8.36 &  8.11 & M3.5V & Rei95 & 17.31$\pm$2.82 & 12 &  \\
LHS 3197   & LP 805-10 & 14.30 & 12.93 & 11.16 & 2 & $R$ & .011 & 14 &  51 &  9.55 &  9.00 &  8.68 & M4.5V & ..... & 14.13$\pm$2.43 & 12 &  \\
LHS 3218   & LP 862-184& 14.18 & 12.93 & 11.28 & 2 & $R$ & .016 & 18 &  77 &  9.78 &  9.27 &  9.00 & M4.0V & Rei95 & 20.86$\pm$3.67 & 12 &  \\
GJ 633     & LHS 3233  & 12.67 & 11.56 & 10.20 & 5 & $V$ & .011 & 20 & 100 &  8.89 &  8.31 &  8.05 & M2.5V & ..... & 19.47$\pm$3.03 & 12 &\tablenotemark{a}\\
LHS 3295   & L 21-3    & 12.18 & 11.02 &  9.53 & 2 & $V$ & .007 & 14 &  68 &  8.09 &  7.52 &  7.30 & M3.0V & ..... & 11.32$\pm$1.75 & 12 &  \\
WT 562     &           & 15.36 & 13.93 & 12.13 & 3 & $I$ & .010 & 18 &  80 & 10.35 &  9.81 &  9.45 & M4.5V & ..... & 16.96$\pm$2.65 & 12 &  \\
LHS 3413   & L 207-41  & 12.68 & 11.44 &  9.88 & 3 & $R$ & .017 & 13 &  69 &  8.32 &  7.70 &  7.46 & M3.5V & Haw96 &  9.85$\pm$1.56 & 12 &  \\
LHS 3443   & L 491-42  & 12.39 & 11.27 &  9.85 & 3 & $V$ & .009 & 14 &  69 &  8.47 &  7.92 &  7.66 & M2.0V & Haw96 & 14.98$\pm$2.30 & 12 &  \\
LHS 3583   & L 23-30   & 11.50 & 10.39 &  9.02 & 3 & $V$ & .014 & 14 &  68 &  7.69 &  7.12 &  6.83 & M2.5V & Haw96 & 10.83$\pm$1.69 & 12 & \tablenotemark{a}\\
APMPM J2127-3844&      & 14.60 & 13.31 & 11.66 & 2 & $R$ & .015 & 12 &  58 & 10.03 &  9.56 &  9.28 & M4.0V & ..... & 21.41$\pm$3.53 & 12 &  \\
WT 795     &           & 14.15 & 12.80 & 11.08 & 3 & $V$ & .015 & 17 &  74 &  9.46 &  8.83 &  8.53 & M4.0V & ..... & 13.11$\pm$2.05 & 12 &  \\
LHS 3719   & L 165-102 & 12.56 & 11.45 & 10.08 & 2 & $V$ & .012 & 14 &  70 &  8.74 &  8.12 &  7.89 & M2.0V & Haw96 & 17.23$\pm$2.69 & 12 &  \\
LHS 3738AB & LP 930-69 & 15.78 & 14.29 & 12.46 & 3 & $R$ & .010 & 24 & 113 & 10.65 & 10.09 &  9.76 & M4.5V & Haw96 & 18.50$\pm$2.96 & 12 & BC \\
LHS 3739   & LP 930-70 & 14.72 & 13.45 & 11.88 & 3 & $R$ & .010 & 24 & 113 & 10.39 &  9.83 &  9.56 & M3.5V & Haw96 & 27.56$\pm$4.50 & 12 & A \\
WT 870     &           & 14.43 & 13.10 & 11.40 & 2 & $R$ & .015 & 15 &  70 &  9.76 &  9.18 &  8.89 & M4.0V & ..... & 15.99$\pm$2.47 & 12 &  \\
LHS 3909   & LP 762-3  & 12.97 & 11.82 & 10.40 & 2 & $R$ & .012 & 12 &  55 &  9.06 &  8.48 &  8.22 & M3.0V & Rei95 & 19.45$\pm$3.09 & 12 &  \\
LHS 3925   & L 359-91  & 13.61 & 12.44 & 10.92 & 2 & $R$ & .008 & 12 &  62 &  9.53 &  8.97 &  8.71 & M3.5V & Haw96 & 21.89$\pm$3.53 & 12 &  \\
LHS 4009AB & G 273-130 & 14.38 & 12.90 & 10.99 & 3 & $R$ & .017 & 17 &  83 &  9.21 &  8.61 &  8.31 & M4.5VJ& ..... &  9.21$\pm$1.51 & 12 &  \\
LHS 4016   & G 275-106 & 12.34 & 11.24 &  9.90 & 2 & $V$ & .016 & 16 &  68 &  8.58 &  8.02 &  7.74 & M2.5V & Rei95 & 17.20$\pm$2.67 & 12 &  \\
LHS 4021   & G 273-147 & 13.44 & 12.19 & 10.59 & 2 & $V$ & .017 & 15 &  60 &  8.94 &  8.39 &  8.04 & M4.0V & Rei95 & 11.84$\pm$1.95 & 12 &  \\
LHS 4058   & G 267-11  & 12.84 & 11.64 & 10.08 & 2 & $V$ & .011 & 16 &  59 &  8.59 &  7.98 &  7.74 & M3.5V & ..... & 12.24$\pm$1.89 & 12 & \tablenotemark{a}\\
\hline
\multicolumn{18}{c}{Beyond 25 pc}\\
\hline
LHS 1561   & G 77-64   & 13.07 & 11.84 & 10.30 & 4 & $V$ & .010 & 13 &  61 &  8.83 &  8.27 &  7.93 & M3.5V & Rei95 & 13.49$\pm$2.14 & 12 &  \\
LHS 1656   & L 178-47  & 13.29 & 12.18 & 10.84 & 2 & $I$ & .009 & 13 &  51 &  9.52 &  8.94 &  8.65 & M2.5V & ..... & 25.68$\pm$4.03 & 12 &  \\
LTT 5066   & LHS 2698  & 14.21 & 13.14 & 11.76 & 2 & $R$ & .010 & 16 &  69 & 10.48 &  9.96 &  9.70 & M3.0V & ..... & 44.38$\pm$7.27 & 12 &  \\
LHS 3080   & G 167-47  & 14.32 & 13.01 & 11.32 & 2 & $R$ & .012 & 17 &  66 &  9.67 &  9.11 &  8.82 & M4.0V & ..... & 15.73$\pm$2.41 & 12 &  \\
LHS 3147   & LP 861-12 & 13.20 & 12.09 & 10.63 & 2 & $R$ & .012 & 18 &  72 &  9.28 &  8.69 &  8.41 & M3.5V & Rei95 & 20.74$\pm$3.23 & 12 &  \\
GJ 762     & LHS 3471  & 12.09 & 11.07 &  9.83 & 2 & $V$ & .008 & 13 &  74 &  8.59 &  8.00 &  7.77 & M2.5V & Haw96 & 20.36$\pm$3.14 & 12 &  \\
LHS 3484   & L 79-24   & 13.88 & 12.70 & 11.19 & 2 & $R$ & .008 & 14 &  65 &  9.79 &  9.22 &  8.98 & M3.5V & Haw96 & 24.66$\pm$3.94 & 12 &  \\
LHS 3836   & L 119-44  & 14.34 & 13.14 & 11.61 & 2 & $R$ & .009 & 11 &  61 & 10.18 &  9.67 &  9.41 & M3.5V & ..... & 29.54$\pm$5.00 & 12 &  \\
\hline
\enddata
\vfil\eject

\tablecomments{Photometry data collected on the sample. All data
collected or calculated by CTIOPI except where otherwise noted.}

\tablenotetext{a}{Astrometric results and relative photometry use new
$V$ filter data.}
\tablenotetext{b}{\citet{1996AJ....112.2799H}}
\tablenotetext{c}{\citet{1995AJ....110.1838R}}
\tablenotetext{d}{\citet{2007AJ....133.2825R} (luminosity class
inferred from the paper, where giants were simply discarded)}
\tablenotetext{e}{\citet{1995AJ....109..797K}}
\end{deluxetable}
}



\thispagestyle{empty}

\voffset000pt{
\begin{deluxetable}{lcclccc}
\setlength{\tabcolsep}{0.04in}
\tablewidth{0pt}
\tabletypesize{\scriptsize}
\tablecaption{Combined system parallaxes\label{tab:weightedmeans}}
\tablehead{\colhead{}              &
	   \colhead{$\pi$}         &
	   \colhead{}              &
	   \colhead{}              &
	   \colhead{$\pi$}         &
	   \colhead{}              &
	   \colhead{Weighted $\pi$}\\

	   \colhead{Name}          &
           \colhead{(mas)}         &
	   \colhead{source}        &
           \colhead{Name}          &
	   \colhead{(mas)}         &
	   \colhead{source}        &
	   \colhead{(mas)}         \\

           \colhead{(1)}           &
           \colhead{(2)}           &
           \colhead{(3)}           &
           \colhead{(4)}           &
           \colhead{(5)}           &
           \colhead{(6)}           &
           \colhead{(7)}           }

\startdata
\hline
LHS 1050  & 85.85$\pm$2.57 & this work & LHS 1050   & 86.60$\pm$13.40 & YPC & 85.88$\pm$2.52 \\
LHS 1749A & 46.14$\pm$1.46 & this work & LHS 1749A  & 45.36$\pm$5.08 & 1.5m & 46.08$\pm$1.40 \\
LHS 2021  & 63.67$\pm$1.15 & this work & LHS 2021   & 59.81$\pm$4.52 & 1.5m & 63.44$\pm$1.11 \\
GJ 438    & 91.70$\pm$2.05 & this work & GJ 438     &118.90$\pm$15.00 & YPC & 92.20$\pm$2.03 \\
LHS 2567  & 46.80$\pm$1.83 & this work & LHS 2568   & 48.55$\pm$1.81 & this work & 47.68$\pm$1.29 \\
LHS 3001  & 57.17$\pm$1.34 & this work & LHS 3002   & 55.64$\pm$1.35 & this work & 56.41$\pm$0.95 \\
GJ 633    & 59.47$\pm$1.19 & this work & GJ 633     &104.00$\pm$13.70 & YPC & 59.80$\pm$1.19 \\
LHS 3739  & 50.97$\pm$1.05 & this work & LHS 3738AB & 50.87$\pm$1.14 & this work & 50.92$\pm$0.77 \\
GJ 762    & 38.72$\pm$1.34 & this work & GJ 762     & 60.30$\pm$14.90 & YPC & 38.89$\pm$1.33 \\
\enddata

\vfil\eject

\tablecomments{`YPC' is \citet{1995gcts.book.....V}; `1.5m' is the
CTIOPI 1.5m program, \citet{2006AJ....132.1234C}.}

\end{deluxetable}
}



\thispagestyle{empty}

\voffset000pt{
\begin{deluxetable}{lcccccc}
\setlength{\tabcolsep}{0.04in}

\tablewidth{0pt}
\tabletypesize{\scriptsize}
\tablecaption{Multiple System Parameters\label{tab:multiples}}
\tablehead{\colhead{Binary}          &
           \colhead{UT Date}         &
           \colhead{Sep.}            &
           \colhead{PA}              &
           \colhead{Period}          &
           \colhead{$\Delta$}        &
           \colhead{}                \\

           \colhead{Name}            &
           \colhead{}                &
           \colhead{(mas)}           &
           \colhead{(deg)}           &
           \colhead{(years)}         &
           \colhead{$mag$}           &
           \colhead{Notes}           \\

           \colhead{1}               &
           \colhead{2}               &
           \colhead{3}               &
           \colhead{4}               &
           \colhead{5}               &
           \colhead{6}               &
           \colhead{7}               }

\startdata
LHS 1582AB & -           & 18.4$\pm$2.8\tablenotemark{a}& -      & 6.35$\pm$0.25  & -    & \\
           &             &                  &                &                &      & \\
LHS 1630AB & -           & -                & -              & $>$9           & -    & (never resolved)\\
           &             &                  &                &                &      & \\
LHS 1749AB & 2002 JAN 31 &  2847.8$\pm$28.5 & 138.2$\pm$0.30 & $>$4           & $\Delta V\sim$2.8\tablenotemark{b} & \\
           & 2004 DEC 28 &  2853.8$\pm$10.9 & 140.0$\pm$0.19 &                &      & \\
LHS 1955AB & 2001 JAN 18 &   808.1$\pm$31.8 & 103.9$\pm$4.40 & $\sim$80       & $\Delta R\sim$0.5\tablenotemark{b} & \\
           & 2009 APR 28 &   948.8:         &  66.5:         &                &      &\tablenotemark{c}\\
LHS 2071AB & -           & 21.2$\pm$4.0\tablenotemark{a}& -  & 16.46$\pm$2.83 &      & period uncertain\\
           &             &                  &                &                &      & \\
LHS 2567/8 & 2000 JAN 27 &  7938.2$\pm$ 3.6 &  61.3$\pm$0.03 & $>$9           & $\Delta V=$1.14 & \\
           & 2009 JAN 13 &  8062.2$\pm$ 1.7 &  61.4$\pm$0.03 &                &      & \\
LHS 3001/2 & 2000 JUL 30 & 12676.9$\pm$ 7.3 &  45.4$\pm$0.02 & $>$9           & $\Delta V=$2.96 & \\
           & 2009 MAR 31 & 12702.6$\pm$ 4.7 &  46.1$\pm$0.02 &                &      & \\
LHS 3739/8 & 1999 OCT 26 &113215.6$\pm$31.4 & 353.7$\pm$0.01 &$>$10           & $\Delta V=$1.06 & (wide) \\
           & 2009 AUG 27 &113115.6$\pm$25.1 & 354.3$\pm$0.01 &                &      & \\
LHS 3738AB & -           & 26.5$\pm$1.8\tablenotemark{a}& -  & 5.85$\pm$0.16  &      & (close) \\
           &             &                  &                &                &      & \\
LHS 4009AB &          -  &     -            &   -            &    -           & -    & (never resolved)\\
           &             &                  &                &                &      & \\

\enddata

\vfil\eject

\tablecomments{All angles and separations measured from the A
component. Separations include a 0.015\% systematic error; angles
include a 0.0083 degree error.}

\tablenotetext{a}{Photocentric semimajor axis}
\tablenotetext{b}{Estimate from relative photometry from relative
parallax reduction}
\tablenotetext{c}{Based on two and one frames, respectively; uses
different centroiding parameters.}

\end{deluxetable}
}



\thispagestyle{empty}

\voffset000pt{
\begin{deluxetable}{lccccccc}
\setlength{\tabcolsep}{0.04in}
\tablewidth{0pt}
\tabletypesize{\scriptsize}
\tablecaption{Preliminary Orbital Elements\label{tab:orbits}}
\tablehead{\colhead{}           &
           \colhead{\emph{P}}   &
           \colhead{\emph{a}\tablenotemark{a}}   &
           \colhead{\emph{i}}   &
           \colhead{Long. Nodes ($\Omega$)}&
           \colhead{\emph{T}}   &
           \colhead{ }          &
	   \colhead{Long. Periastron ($\omega$)}\\

           \colhead{Designation}&
           \colhead{(yr)}       &
           \colhead{(arcsec)}   &
           \colhead{(deg)}      &
           \colhead{(deg)}      &
           \colhead{(yr)}       &
           \colhead{\emph{e}}   &
	   \colhead{(deg)}      \\

           \colhead{1}          &
           \colhead{2}          &
           \colhead{3}          &
           \colhead{4}          &
           \colhead{5}          &
           \colhead{6}          &
           \colhead{7}          &
           \colhead{8}          }

\startdata
LHS 1582AB &  6.4$\pm$0.2 & 0.018$\pm$0.003 & 140.8$\pm$14.1 & 279.7$\pm$24.0 & 2002.2$\pm$0.3 & 0.19$\pm$0.06 & 258.4$\pm$23.0 \\
LHS 2071AB &16.4$\pm$2.8\tablenotemark{b}&0.021$\pm$0.004&090.9$\pm$2.6&219.1$\pm$3.6&2005.9$\pm$1.0&0.31$\pm$0.12&33.2$\pm$26.23\\
LHS 3738AB &  5.8$\pm$0.2 & 0.027$\pm$0.002 & 118.9$\pm$5.16 & 315.9$\pm$ 5.5 & 2006.9$\pm$1.3 & 0.04$\pm$0.04 &  26.7$\pm$82.7 \\
\enddata

\vfil\eject

\tablenotetext{a}{Photocentric semimajor axis}
\tablenotetext{b}{Highly uncertain}

\end{deluxetable}
}



\thispagestyle{empty}

\voffset000pt{
\begin{deluxetable}{llll}
\setlength{\tabcolsep}{0.04in}
\tablewidth{0pt}
\tabletypesize{\scriptsize}
\tablecaption{Comparison of spectral types\label{tab:spectra}}
\tablehead{\colhead{}              &
	   \colhead{SpType}        &
	   \colhead{SpType}        &
	   \colhead{}              \\

	   \colhead{Name}          &
           \colhead{RECONS}        &
	   \colhead{PMSU}          &
	   \colhead{Ref}           \\

           \colhead{(1)}           &
           \colhead{(2)}           &
           \colhead{(3)}           &
           \colhead{(4)}           }

\startdata
\hline
LHS1050   & M3.0V & M3.0 & 1\\
LHS1630AB & M3.5VJ& M3.5 & 2\\
LHS1767   & M3.0V & M3.5 & 2\\
LHS2071AB & M4.0VJ& M4   & 2\\
LHS2836   & M4.0V & M4   & 2\\
LHS3197   & M4.5V & M4.5 & 2\\
LHS3295   & M3.0V & M3.5 & 2\\
LHS4009AB & M4.5VJ& M5   & 2\\
LHS4058   & M3.5V & M3.0 & 2\\
LHS3080   & M4.0V & M4.5 & 1\\
LHS3836   & M3.5V & M3.5 & 2\\
\enddata

\vfil\eject

\tablecomments{Comparison of our spectral types versus PMSU for all
overlapping systems.  1.) \citet{1995AJ....110.1838R} 2.)
\citet{1996AJ....112.2799H} }

\end{deluxetable}
}

\end{document}